\documentclass[12pt]{article}
\usepackage{amsmath,amsfonts,amssymb}
\usepackage{citesort}
\usepackage{epsfig}

\def\tev{~{\rm TeV}}
\def\gev{~{\rm GeV}}
\def\bit{\begin{itemize}}
\def\eit{\end{itemize}}
\def\ben{\begin{enumerate}}
\def\een{\end{enumerate}}
\def\bed{\begin{description}}
\def\eed{\end{description}}
\def\susy{{\sc Susy}}
\def\Susy{{\sc Susy}}
\def\SUSY{{\sc Susy}}
\def\b{\beta}

\def\k{\kappa}
\def\l{\lambda}
\def\t{\theta}
\def\q{\quad}
\def\qq{\qquad}
\def\wt{\widetilde}
\def\half{\frac{1}{2}\,}
\def\third{\frac{1}{3}\,}
\def\quart{\frac{1}{4}\,}
\def\R{ {\rm R \kern -.31cm I \kern .15cm}}
\def\C{ {\rm C \kern -.15cm \vrule width.5pt \kern .12cm}}
\def\Z{ {\rm Z \kern -.27cm \angle \kern .02cm}}
\def\N{ {\rm N \kern -.26cm \vrule width.4pt \kern .10cm}}
\def\1{{\rm 1\mskip-4.5mu l} }
\def\lsim{\raise0.3ex\hbox{$<$\kern-0.75em\raise-1.1ex\hbox{$\sim$}}}
\def\gsim{\raise0.3ex\hbox{$>$\kern-0.75em\raise-1.1ex\hbox{$\sim$}}}
\def\noi{\noindent}

\def\beq{\begin{equation}}   
\def\eeq{\end{equation}}
\def\bea{\begin{eqnarray}}  
\def\eea{\end{eqnarray}}
\newcommand{\ba}{\begin{array}}
\newcommand{\ea}{\end{array}}
\def\nn{\nonumber}
\def\noi{\noindent}
\def\beeq{\begin{eqnarray}} \def\eeeq{\end{eqnarray}}
\newcommand\mysection{\setcounter{equation}{0}\section}
\renewcommand{\theequation}{\thesection.\arabic{equation}}
\newcounter{hran} \renewcommand{\thehran}{\thesection.\arabic{hran}}

\def\bmini{\setcounter{hran}{\value{equation}}
   \refstepcounter{hran}\setcounter{equation}{0}
   \renewcommand{\theequation}{\thehran\alph{equation}}\begin{eqnarray}}

\def\bminiG#1{\setcounter{hran}{\value{equation}}
\refstepcounter{hran}\setcounter{equation}{-1}
\renewcommand{\theequation}{\thehran\alph{equation}}
\refstepcounter{equation}\label{#1}\begin{eqnarray}}

\def\emini{\end{eqnarray}\relax\setcounter{equation}{\value{hran}}\renewcommand{\theequation}{\thesection.\arabic{equation}}}

\begin{document}
\centerline{\Large\bf NMHDECAY: A Fortran Code for the Higgs Masses,}
\par \vskip 3 truemm
\centerline{\Large\bf  Couplings and Decay Widths in the NMSSM}
\vskip 1 truecm

\begin{center}
{\bf Ulrich Ellwanger}\footnote{E-mail :
Ulrich.Ellwanger@th.u-psud.fr}\par \vskip 5 truemm

Laboratoire de Physique Th\'eorique\footnote{Unit\'e
Mixte de Recherche - CNRS - UMR 8627}\par
  Universit\'e de Paris XI, B\^atiment
210, F-91405 Orsay Cedex, France\par \vskip 5 truemm

{\bf John F. Gunion}\footnote{E-mail: gunion@physics.ucdavis.edu}\par \vskip 5 truemm

Department of Physics \par University of California at Davis \par
Davis, CA 95616, U.S.A.\par

\vskip 5 truemm
{\bf Cyril Hugonie}\footnote{E-mail :
hugonie@ific.uv.es}\par \vskip 5 truemm

Instituto de F\'\i sica Corpuscular -- CSIC/Universitat de Val\`encia
\par
Edificio Institutos de Investigaci\'on, Apartado de Correos 22085,\par
E-46071 Val\`encia, Spain

\end{center}
\vskip 1 truecm

\begin{abstract} The Fortran code NMHDECAY computes the masses,
couplings and decay widths of all Higgs bosons of the NMSSM in terms of
its parameters at the electroweak (or \susy\ breaking) scale: the Yukawa
couplings $\lambda$ and $\kappa$, the soft trilinear terms $A_{\lambda}$ and
$A_\kappa$, and $\tan \beta$ and $\mu_\mathrm{eff} = \lambda \left< S
\right>$. The computation of the spectrum includes leading two loop
terms, electroweak corrections and propagator corrections. The
computation of the decay widths is carried out as in HDECAY, but
(for the moment) without three body decays. Each point in parameter space is
checked against negative Higgs bosons searches at LEP, including
unconventional channels relevant for the NMSSM. One version of the
program uses generalized SLHA conventions for input and output.
\end{abstract}

\vfill
\noi LPT Orsay 04-32\par
\noi UCD-04-22\par
\noi IFIC/04-33\par
\noi June 2004 \par

\newpage \pagestyle{plain} \baselineskip 20pt

\mysection{Introduction}

\hspace*{\parindent}
The Next to Minimal Supersymmetric Standard Model (NMSSM
\cite{allg,bast,abel1,abel2,radcor1,radcor2,higsec1,higsec2,higsec3,yeg,higrad2}) 
provides a very elegant solution to the $\mu$ problem of the MSSM 
via the introduction of a singlet superfield $\widehat{S}$. For the
simplest possible scale invariant form of the superpotential,
the scalar component of $\widehat{S}$ acquires naturally a vacuum
expectation value of 
the order of the \susy\ breaking scale, giving rise to a value of $\mu$ 
of order the electroweak scale. 
The NMSSM is actually the simplest
supersymmetric extension of the standard model in which the electroweak
scale originates from the \susy\ breaking scale only. \par

In addition, the NMSSM renders the ``little fine tuning problem'' of
the MSSM, originating from the non-observation of a neutral CP-even
Higgs boson at LEP II, less severe \cite{bast}.\par

A possible cosmological domain wall problem \cite{abel1} can be avoided
by introducing suitable non-renormalizable operators \cite{abel2} that
do not generate dangerously large singlet tadpole diagrams
\cite{tadp}.\par

Hence, the phenomenology of the NMSSM deserves to be studied at least as
fully and precisely as that of the MSSM. Its particle content differs from
the MSSM by the addition of one  CP-even and one CP-odd state in
the neutral Higgs sector (assuming CP conservation), 
and one additional neutralino. Thus,
the physics of the Higgs bosons -- masses, couplings and branching
ratios \cite{allg,radcor1,radcor2,higsec1,higsec2,higsec3,yeg,higrad2} can differ significantly from the MSSM. The purpose
of the Fortran code NMHDECAY (Non Minimal Higgs Decays), that
accompanies the present paper, is an accurate computation of these
properties of the Higgs bosons in the NMSSM in terms of the parameters
in the Lagrangian. As its name suggests, the Fortran code uses to some
extent -- for MSSM-like processes -- parts of the code HDECAY that is
applicable to the Higgs sector of the MSSM \cite{hdecay}.\par

In the present paper we define the NMSSM in terms of its parameters at
the \susy\ breaking scale. No assumption on the soft terms (like
universal soft terms at a GUT scale) are made. The parameters in the
Higgs sector are chosen as follows:
\bed
\item{a)} Apart from the usual quark and lepton Yukawa couplings, 
the scale invariant superpotential is
\beq \label{1.1} \lambda \ \widehat{S} \widehat{H}_u
\widehat{H}_d + \frac{\kappa}{3} \ \widehat{S}^3 \eeq
\noi depending on two dimensionless couplings $\lambda$, $\kappa$ beyond the
MSSM.
(Hatted capital letters denote superfields, and unhatted capital
letters will denote their scalar components).

\item{b)} The associated trilinear soft terms are
\beq \label{1.2} \lambda A_{\lambda} S H_u H_d + \frac{\kappa}
{3} A_\kappa S^3\,. \eeq

\item{c)} The final two input parameters are
\beq \label{1.3} \tan \beta =\ \left< H_u \right>/ \left< H_d \right>\ ,
 \ \mu_\mathrm{eff} = \lambda \left< S \right>\ . \eeq
These, along with $M_Z$, can be viewed as determining the
three \susy\ breaking masses squared for $H_u$, $H_d$
and $S$ through the three minimization equations of the
scalar potential.
\eed
Thus, as
compared to two independent parameters in the Higgs sector of the
MSSM (often chosen as $\tan \beta$ and $M_A$), the Higgs sector of
the NMSSM is described by the six parameters
\beq \label{6param}
\lambda\ , \ \kappa\ , \ A_{\lambda} \ , \ A_{\kappa}, \ \tan \beta\ ,
\ \mu_\mathrm{eff}\ .
  \eeq
We will choose sign conventions for the fields such that $\lambda$ and
$\tan\beta$ are positive, while $\kappa$, $A_\lambda$, $A_{\kappa}$ and
$\mu_{\mathrm{eff}}$ should be allowed to have either sign.  For any
choice of these parameters -- as well as of the values for the gaugino
masses  and of the soft terms related to the squarks and sleptons that
contribute to the radiative corrections in the Higgs sector and to the
Higgs decay widths -- NMHDECAY performs the following tasks:
\ben
\item It computes the masses and couplings of all physical states in
the Higgs, chargino and neutralino sectors. Error messages are produced
if a Higgs or squark mass squared is negative.

\item It computes the branching ratios into two particle final states
(including char\-ginos and neutralinos --- decays to squarks and
sleptons will be implemented in a later release) of all Higgs
particles.

\item It checks whether the Higgs masses and couplings violate any
bounds from negative Higgs searches at LEP, including many quite
unconventional channels that are relevant for the NMSSM Higgs sector.
It also checks the bound on the invisible $Z$ width (possibly violated
for light neutralinos). In addition, NMHDECAY checks the bounds on the
lightest chargino and on neutralino pair production. Corresponding
warnings are produced in case any of these phenomenological constraints
are violated.

\item It checks whether the running Yukawa couplings encounter a Landau
singularity below the GUT scale. A warning is
produced if this happens.

\item Finally, NMHDECAY checks whether the physical minimum (with all
vevs  non-zero) of the scalar potential is deeper than the local
unphysical minima with vanishing $\left<H_u \right>$ or $\left<H_d
\right>$. If this is not the case, a warning is produced.

\een

The web sites \par

\hspace*{.5in}{\tt{http://www.th.u-psud.fr/NMHDECAY/nmhdecay.html}}

\hspace*{.5in}{\tt{http://higgs.ucdavis.edu/nmhdecay/nmhdecay.html}}

\noi provide links for downloading two different Fortran codes, 
along with descriptions, input
files, and the data files that are required to apply the various LEP constraints.
\par

The Fortran code NMHDECAY\_SLHA.f reads from input files, and produces
output files, that are as close as possible to the \Susy\ Les Houches
Accord (SLHA) conventions \cite{slha}. Some generalizations of these
conventions -- including proposals for PDG numbers -- have been
necessary, however, in order to denote the NMSSM input parameters
(\ref{6param}) and the additional particles in the Higgs and neutralinos
sectors.\par

The Fortran code NMHDECAY\_SCAN.f reads from ``private'' input files
(samples are provided) that are constructed so as to scan over the NMSSM
input parameters (\ref{6param}). Here the output is either ``long''
(easily human readable, if the number of points in parameter space is
not too large), or ``short'', i.e. simple rows of numbers per point in
parameter space, that should be edited according to the user's
needs.\par

Note that the ``long'' output also gives the reduced couplings of all
neutral Higgs bosons to gauge bosons (CV), up type quarks (CU), down
type quarks (CD), two gluons (CG) and two photons (CGA) (all relative
to a Standard Model Higgs boson of the same mass). Using these, it is
easy to  compute the NMSSM Higgs production cross sections at colliders
by rescaling those for the SM Higgs boson. \par

The outline of the paper is as follows: In section~\ref{sec:radcor}, 
we describe in detail the accuracy with which the Higgs masses and
mixing matrices are computed. In section~\ref{sec:higgsdecays}, we
describe the Higgs decay channels and  corresponding accuracies that
are used for the computation of the widths and branching ratios. In
section~\ref{sec:constraints}, we describe the various 
phenomenological constraints that can be applied to the model.
Section~\ref{sec:use}  contains a user's guide for the Fortran codes,
as well as our generalizations of the SLHA conventions. In
section~\ref{sec:results}, we show some results for branching ratios
produced with the present code, emphasizing decay channels that are
atypical of the MSSM, such as the lightest CP-even state $h_1$ decaying
to 2 lighter CP-odd states $a_1 a_1$. Section~\ref{sec:conclusions}
contains our conclusions.  Appendix~A contains a summary of our
conventions, appendix~B a summary of the Feynman rules, and appendix~C
the details of the radiative corrections to the higgs mass matrices.

\mysection{Radiative Corrections in the Higgs Sector}
\label{sec:radcor}

\hspace*{\parindent}
Our convention for the superpotential, the soft  terms, the resulting
tree level potential and tree level mass matrices are given in appendix
A. In the present section, we describe the accuracy with which
radiative corrections are computed for the Higgs sector.\par

First, we assume that the Yukawa couplings and soft terms are defined
at a \susy\ breaking scale $Q = M_\mathrm{SUSY}$, which corresponds to
an average of the squark masses. Quantum fluctuations at scales $> Q$
are assumed to be integrated out, which corresponds to the standard RG
evolution of the parameters from a fundamental scale like
$M_\mathrm{GUT}$ down to $M_\mathrm{SUSY}$. The effective Lagrangian at
the scale $Q$ can be assumed to be of the standard supersymmetric form
plus soft terms.\par

We are interested in the full effective action
\beq \label{2.1} \Gamma_\mathrm{eff} = \sum_\mathrm{i} Z_i \ D_{\mu}
H_i  D^{\mu} H_i - V_\mathrm{eff}(H_i) \ , \eeq

\noi that is obtained from the effective Lagrangian at the scale $Q$ by
adding all quantum effects arising from fluctuations at scales $\lsim\
Q$. (Here, $H_i$ denotes all the Higgs fields, $H_u$, $H_d$ and $S$.)
These quantum effects can be classified according to powers of the
various couplings, and powers of potentially large logarithms. Explicit
formulas for the radiative corrections to $V_{eff}$, $Z_i$ and hence to
the Higgs mass matrices are given in appendix~C, subsequently we
describe only the orders to which radiative corrections are taken into
account. \par

Let us start with the corrections to the effective potential. It is
somewhat more convenient, however, to classify the corrections to the
(lightest) scalar Higgs mass, which is essentially the second
derivative of the effective potential. At tree level this mass squared
is bounded by
\beq \label{2.2} m_1^2 < M_Z^2 \left ( \cos^2 2 \beta + 
\frac{2 \lambda^2 \sin^2 2 \beta}{g_1^2 + g_2^2}\right ) \ , \eeq

\noi which is of the order
\beq \label{2.3} m_1^2 \sim {\cal O}(g^2v^2) + {\cal O}(\lambda^2 v^2)
\eeq

\noi where $g$ denotes the electroweak gauge couplings, and $v$ the
magnitude of the Higgs vacuum expectation values (we do not distinguish
between large and small $\tan \beta$ in this section).\par

The dominant one-loop corrections to $m_1^2$ originate from top, stop,
bottom and sbottom loops. The corresponding corrections $\delta m_1^2$
to $m_1^2$ are of order
\bea \label{2.4}
&&\delta_\mathrm{top} m_1^2 \sim h_t^4 \ v^2 \ln \left (
Q^2/m_\mathrm{top}^2 \right )  \ , \nn \\
&&\delta_\mathrm{stop} m_1^2 \sim h_t^4 \ v^2 \hskip 3,5cm  ({\rm for}\
Q^2 \sim  m_\mathrm{stop}^2 ) \ , \nn \\
&&\delta_\mathrm{bottom} m_1^2 \sim h_b^4 \ v^2 \ln \left (
Q^2/m_\mathrm{bottom}^2  \right ) \ , \nn \\
&&\delta_\mathrm{sbottom} m_1^2 \sim h_b^4 \ v^2  \hskip 3cm ( {\rm
for}\ Q^2 \sim m_\mathrm{sbottom}^2 ) \ .
\eea

\noi We have included these contributions \cite{radcor1,radcor2,yeg}
 exactly, without
expanding in large logarithms or squark mass splittings.\par

The dominant two loop corrections are of the form \cite{yeg,higrad2}
\beq \label{2.5}
\delta^{(2)} m_1^2 \sim h_t^6 v^2 \ln^2 \left ( Q^2/m_\mathrm{top}^2
\right )  + h_t^4 \alpha_s v^2 \ln^2 \left ( Q^2/m_\mathrm{top}^2
\right ) \ .
\eeq

\noi Here, we only include the leading double logarithms, i.e. we
neglect single logs as well as terms involving $h_b^2$. \par

Coming back to the one-loop order, we focus on including all
contributions of the order $g^4 \times$ large logarithms. These can be
obtained by RG methods \cite{haber}, taking care of the fact that the
electroweak couplings $g$ are defined at the scale $M_Z$ and not at the
scale $M_\mathrm{SUSY}$. We have included all these contributions as
given in \cite{haber}, distinguishing large logarithms $\ln
(Q^2/M_Z^2)$ from $\ln(M_A^2/ M_Z^2)$ and $\ln (\mu_i^2/M_Z^2)$ ($\mu_i
= M_1, M_2$ or $\mu_\mathrm{eff}$, where $M_i$ are the electroweak
gaugino masses). We have not (yet) included terms of the order
$g^2\lambda^2 \times$ large logarithms, nor terms of the order $g^4
\times$ finite. \par

This concludes our contributions to the effective potential. Next, we
outline the contributions to the wave function normalization constants
$Z_i$ in (\ref{2.1}). If evaluated for external momenta of ${\cal
O}(m_\mathrm{top})$ (the order of the Higgs masses), top and bottom
quark loops yield contributions to $\delta Z_i$ (relative to 1) of the
form
\beq \label{2.6}
\delta Z_1 \sim h_t^2 \ \ln \left ( Q^2/m_\mathrm{top}^2\right ) \ , \ 
\delta Z_2 \sim h_b^2 \ \ln \left ( Q^2/m_\mathrm{top}^2\right ) \ .
\eeq

\noi After rescaling the Higgs fields so that their kinetic energies
are canonically normalized, these effects generate contributions
$\delta_Z m_1^2$ which take the form
\beq \label{2.7}
\delta_Z m_1^2 \sim  g^2 h_{t/b}^2 v^2 \ln \left ( 
Q^2/m_\mathrm{top}^2\right ) + \lambda^2 h_{t/b}^2 v^2 \ln \left ( 
Q^2/m_\mathrm{top}^2\right ) \ .
\eeq

However, after diagonalization of the Higgs mass matrices, one may find
eigenstates with masses $m_H$ substantially larger than
$m_\mathrm{top}$. Then the $Z$ factors should be evaluated at external
momenta of ${\cal O}(m_H)$, i.e. at the pole of the propagators. Hence
for $m_H \gg m_\mathrm{top}$ the logarithms $\ln
(Q^2/m_\mathrm{top}^2)$ in (\ref{2.6}) should be replaced by $\ln
(Q^2/m_H^2)$, with coefficients depending on Higgs mixing matrices.\par

All this is done in the program, with the net result that all
potentially large logarithms with coefficients $g^2h_{t/b}^2$ or
$\lambda^2 h_{t/b}^2$ have the correct arguments. (We neglect, however,
contributions of ${\cal O}(\ln(m_H^2/m_{top}^2))$ in the off diagonal 
Higgs matrix elements, which affects only the couplings of very heavy
Higgses.)
Since we have
neglected terms of ${\cal O}(h_{t/b}^2\times finite)$ in the $Z$ factors
and electroweak corrections without large logarithms, the dominant
sources of uncertainty in $\delta m^2$ are contributions of 
${\cal O}(g^4 v^2\times finite)$,
${\cal O}(g^2 h_{t/b}^2 v^2\times finite)$ and 
${\cal O}(\lambda^2 h_{t/b}^2 v^2\times finite)$ as well as subdominant
(single log) two loop corrections.
\par

We have checked that in the end, after expanding the $Z$ factors
appropriately, all $Q^2$ dependence in the one-loop contributions can
be reabsorbed into the running of the input parameters that are defined
at the scale $Q^2$. An exception is the $Q^2$ dependence of terms $\sim
g^4$, since we have not included the squark loop contributions $\sim
g^4$ (from squark couplings to Higgs bosons arising from the
$D$-terms). These would give contributions to $m_1^2$ of the order
\beq \label{2.8}
\delta_D m_1^2 \sim g^4 \ln \left ( Q^2/m_\mathrm{squark}^2 \right )\ ,
\eeq

\noi which is relatively small for $Q^2 \sim m_\mathrm{squark}^2$.\par

Since we have not included contributions of the orders $(g^4,\ g^2
h_{t,b}^2) \times$ finite, we are actually not sensitive to the scheme
in which the electroweak couplings are defined. Since in the
computation of the decay widths we use the physical masses $M_W$ and
$M_Z$ in order to parameterize some couplings, the most useful scheme
here is the on-shell scheme, where the three ``unphysical'' electroweak
parameters $v^2 = \left< H_u^2 + H_d^2 \right>$, $g_1^2$ and $g_2^2$
are defined in terms of $G_F$, $M_W$ and $M_Z$ as
\bea
\label{2.9}
v^2 &=& \frac{1}{2 \sqrt{2} \ G_F} \ , \nn \\
g_2^2 &=& 2 M_W^2/v^2 \ , \nn \\
g_1^2 + g_2^2 &=& 2 M_Z^2/v^2 \ .
\eea

Finally we remark that we have applied the above corrections
consistently to the complete Higgs sector, i.e. the CP-even, CP-odd and
charged Higgs states. One finds that one can reabsorb several of the
radiative corrections into a redefinition of the input parameter
$A_{\lambda}$ \cite{radcor1}, after which the CP-odd mass matrix assumes
its tree level form, up to the required rescalings of the vevs
by the $Z$ factors. This phenomenon is familiar from the MSSM, where
$M_A$ is thus a convenient input parameter, and the required rescalings
are often absorbed into a scale dependent value of $\tan \beta$. \par

We have checked that in the MSSM limit of the NMSSM ($\lambda , \kappa
\ll 1$), after comparing for the same values of $\tan \beta$ and $M_A$
(and the remaining soft terms), the mass of the lightest CP-even state
agrees with the one computed in HDECAY \cite{hdecay} (models 1
\cite{car1} or 2 \cite{car2}) to better than 2 GeV for moderate mixing
($A_\mathrm{top} \ \lsim$~1~TeV), and to better than 3 GeV for maximal
mixing. This coincides with the expected theoretical error on this mass
given the orders of uncertainty discussed above.\par

A more detailed discussion of the radiative corrections to the NMSSM
Higgs sector will appear in \cite{toapp}.

\mysection{Higgs Decays}
\label{sec:higgsdecays}

\hspace*{\parindent}
In this section we describe the decay modes that
are included in the Fortran code. Most of the corresponding code is
extracted from HDECAY \cite{hdecay}. In what follows, $H$ denotes any
of the 3 CP-even or 2 CP-odd scalars, or the charged Higgs boson. \par

a) $H \to$ gluons: We take into account charm, bottom and top quark
loops; the lowest order contribution to the decay width is given in
\cite{ellisnan}. QCD radiative corrections \cite{gluqcd} are included at
the same level as in HDECAY. Squark loops are not (yet) taken into
account.\par

b) $H \to$ leptons and quarks: QCD corrections are included as in
HDECAY \cite{quqcd}, but we use always the two loop formula for the
relation between the pole quark mass $m_\mathrm{pole}$ and the running
$\overline{MS}$ quark mass $\overline{m}(m_\mathrm{pole})$:

\beq \label{3.1} \overline{m}_i(m_\mathrm{pole}) =
\frac{m_\mathrm{pole}}{1 + \frac{4 \alpha_s (m_\mathrm{pole})}{3 \pi} +
K_i \left ( \frac{\alpha_s (m_\mathrm{pole}) }{\pi} \right )^2} \eeq

\noi with $K_\mathrm{bottom} \sim 12.4$, $K_\mathrm{top} \sim 11$. The
scale dependence of the running $s$ and $c$ quark masses is computed as
in HDECAY, but for the running $b$ and $t$ quark masses we use running
Yukawa couplings that solve the one-loop RG equations exactly in
$\alpha_s$, improved by corrections to first order in the Yukawa
couplings themselves (see appendix C).

Since electroweak corrections to decays into leptons and quarks are
small \cite{hdecay}, they are neglected. Below-threshold three-body
decays $H \to t\overline{t} \to t \overline{b}W^-$ are not (yet)
taken into account. However, contributions to the decay rates into
$b\overline{b}\ (c\overline{c})$ from $H \to gg^* \to
gb\overline{b}\ (c\overline{c})$ are included as in HDECAY (option NF-GG
$ = 3$, that can easily be modified by re-editing the
line NFGG $= 3$ at the beginning of the subroutine DECAY). \Susy\ loop
corrections are not (yet) included. (These could be important for light
squarks and/or large $\tan{\beta}$.) In the case of a decaying charged
Higgs boson, we must input the CKM matrix elements $V_{us}$, $V_{ub}$ and
$V_{cb}$. 
\par

c) $H \to WW/ZZ$: We treat these decays as in HDECAY \cite{hww},
but without the double off-shell option.\par

d) $H \to \gamma \gamma /Z\gamma$: We take into account loops of charm,
bottom and top quarks, $\tau$ leptons, $W$ bosons, charginos and
charged Higgs bosons \cite{hgaga}. Loops of squarks and sleptons are
not (yet) considered.\par

e) $H \to HH$: All kinematically possible combinations ($H =$ CP-even,
CP-odd or charged) are considered, with (Higgs)$^3$-couplings computed
from the NMSSM potential (see F. Franke and H. Fraas in
ref.~\cite{allg} and appendix B). The dominant (leading log) radiative
corrections from top-quark and bottom-quark loops to these couplings
are included in the Fortran code, the corresponding formulas will
appear in \cite{toapp}. Decays into three particle final states are not
(yet) taken into account.\par

f) $H \to HZ/HW$: All kinematically possible combinations ($H =$
CP-even, CP-odd or charged) are considered \cite{hhz}. Below threshold
decays into three particle final states are not (yet) taken into
account.\par

g) $H \to$ charginos/neutralinos: First, the $2 \times 2$ chargino and
$5 \times 5$ neutralino mass matrices are diagonalized in the
subroutines CHARGINO and NEUTRALINO, and the decays into all possible
two body final states \cite{hcc} are included in the program. \par

Decays into sleptons or squarks are not (yet) considered. Note that
relatively light squarks -- even if too heavy to be produced in Higgs
decays -- could nevertheless affect notably the processes b) and d)
above through squark loops. For this reason the present version of
NMHDECAY should not be applied to scenarios with light squarks.

\mysection{Constraints on the Parameter Space}
\label{sec:constraints}

\hspace*{\parindent}
Once the spectrum and decay widths of the Higgses are computed, we
check some experimental and theoretical constraints for each point in
parameter space:

First, we require that there is no Landau pole for the Yukawa couplings
$\lambda, \kappa, h_\mathrm{top}$ and $h_\mathrm{bot}$ below the GUT
scale.  For this, we integrate the two loop renormalization group
equations for the Yukawa and gauge couplings from the \SUSY\ scale up
to the point where the gauge couplings $g_1$ and $g_2$ unify. At this
point we check that none of the couplings is too large (i.e. exceeds
$\sqrt{4\pi}$). (For simplicity we assume a universal \SUSY\ threshold
at $M_{susy} \sim Q$.)

We also check that there is no deeper minimum of the scalar potential
with vanishing $\left<H_u \right>$ or $\left<H_d \right>$.

Finally we check all available experimental constraints from LEP:

\noi 1) In the neutralino sector, we check that the lightest neutralino
does not contribute too much to the invisible $Z$ width ($\Gamma(Z \to
\chi^0_1 \chi^0_1) < 1.76$ MeV \cite{LEPneu}) if $m_{\chi^0_1} <
M_Z/2$, and that $\sigma(e^+e^- \to \chi^0_1 \chi^0_i) < 10^{-2}~{\rm
pb}$ if $m_{\chi^0_1} + m_{\chi^0_i} < 209$ GeV ($i>1$) and
$\sigma(e^+e^- \to \chi^0_i \chi^0_j) < 10^{-1}~{\rm pb}$ if
$m_{\chi^0_i} + m_{\chi^0_j} < 209$ GeV ($i,j>1$) \cite{LEPneu2}. In
the chargino sector, we check that the lightest chargino is not too
light ($m_{\chi^+_1} < 103.5$ GeV \cite{LEPchar}).
 
\noi 2) In the charged Higgs sector, we check the bound $m_{H^+} >
78.6$ GeV \cite{LEPHC}.

\noi 3) In the neutral Higgs sector, we check the constraints on the
production rates (reduced couplings) $\times$ branching ratios versus
the masses, for all of the CP-even states $h$ and CP-odd states $a$, in
the following channels studied at LEP:
\bit
\item
$e^+e^- \to hZ$, with $h\to b\bar{b}$ and $h\to \tau^+\tau^-$ (from the
LEP Higgs working group results \cite{LHWGSM});

\item $e^+e^- \to hZ$, with $h$ decaying to two jets. For this, we
combined the low mass range results from OPAL at LEP2 \cite{OPAL1} with
the higher mass range obtained at LEP2 by the LEP Higgs Working group
\cite{LHWGjj};
         
\item $e^+e^- \to hZ$, with $h\to \gamma\gamma$ (from the LEP
Higgs working group results \cite{LHWGgg});
                                     
\item $e^+e^- \to hZ$, with $h$ decaying invisibly ({\it i.e.} into two
neutralinos). For this, we combined the low mass range results from
ALEPH at LEP1 \cite{ALEPH} with the higher mass range obtained at LEP2
by the LEP Higgs Working group \cite{LHWGinv}.
                                  
\item $e^+e^- \to hZ$, independent of the $h$ decay mode, looking for a
peak of the $M_X$ recoil mass distribution in $e^+e^-\to XZ$. For this,
we combined the low mass range results from ALEPH at LEP1 \cite{ALEPH}
with the higher mass range obtained at LEP2 by OPAL \cite{OPAL2}.
              
\item For the associated production mode $e^+e^- \to ha$ we used the
DELPHI results \cite{DELPHI} for the various final states: $ha \to 4b$,
$ha \to 4\tau$ and $ha \to aaa \to 6b$. The latter decay channel plays
an important role in the NMSSM, where the lightest CP-even Higgs can
decay mainly into two light CP-odd Higgses over large areas of the
parameter space, as we will see in section~\ref{sec:results}. 
(In the context of the CP-conserving MSSM, this is possible only for
very special parameter choices.) DELPHI also studied the channel $hZ
\to aaZ \to 4b\ +\ 2$jets \cite{DELPHI}; their limits have also been
implemented in our phenomenological constraints.

\item The channels $e^+e^- \to hZ \to aaZ$ with $aa \to 4$jets, 2jets +
$cc$,  2jets +  $\tau\tau$, 4$\tau$'s, $cccc$, $\tau\tau + cc$ have
been studied by OPAL \cite{Opal3} (see their Fig. 7 ---
we employ numerical files for the contours
provided by the authors). 
These limits have also been included. 

\eit

All these constraints are contained in the files located in the
directory LEPCON downloadable together with the Fortran codes.
The file names are constructed so that they can be identified with
the above delineated constraints.
(The current numerical limits contained in these files
can possibly be improved with the help of
further dedicated analysis of existing data.)

\mysection{How to use NMHDECAY}
\label{sec:use}

NMHDECAY exists in two versions:\par

\ben
\item 
 NMHDECAY\_SLHA uses an input file and produces output files
that are suitable generalizations of the SLHA conventions \cite{slha}.
It is designed for studying the properties of one user-defined point in
parameter space.

\item  NMHDECAY\_SCAN uses input and output files that can
be modified by the user so as to
scan over parts of, or all of the NMSSM parameters defined
in eq.~(\ref{6param}).
\een
Both programs have the
common feature that they are based on one single
Fortran code (NMHDECAY\_SLHA.f or NMHDECAY\_SCAN.f) that does not need
to be linked with any other code. However, both programs need data
files in order to check against negative Higgs searches at LEP in the
numerous channels discussed in section 4. These data files are
available in the directory LEPCON, that can be downloaded from the
NMHDECAY home page. Note that the directory LEPCON has to be situated
in the same directory that contains the executable NMHDECAY code.
We now discuss the particular features of the two programs.

\subsection{NMHDECAY\_SLHA}

NMHDECAY\_SLHA uses an input file {\tt slhainp.dat}, a version of which
is  downloaded automatically with the Fortran code. This sample file
appears in Table~\ref{slhainp}. Several comments on its contents are in
order.\par

\noi a) ``BLOCK MODSEL'' contains the entry 3 (corresponding to
the choice of the particle
content) with switch 1, as attributed to the NMSSM in
\cite{slha}.\par

\noi b) ``BLOCK SMINPUTS'' contains important Standard Model
parameters.
\ben
\item
First, there is the inverse electromagnetic coupling constant at the
scale 0, which is that required for the computations of the decay widths
into two (on-shell) photons.

\item Second, various Higgs couplings are defined in terms of $M_Z$,
$\alpha_s(M_Z)$,  $M_W$ and $G_F$, which means an on shell scheme is
implicitly being used in order to define the electroweak parameters (cf.
the corresponding discussion in section 2).

$M_Z$, $\alpha_s(M_Z)$ and $G_F$ are input as part of ``BLOCK
SMINPUTS'', whereas  the numerical value of $M_W = 80.42$~GeV is defined
in the subroutine INPUT, that reads also the input file {\tt
slhainp.dat}. (Part of the corresponding code is copied from {\tt
pyslha.f}, with thanks to P. Skands.) 
\item
Third, as part of this block we input the running $b$ quark mass
$m_b(m_b)$, the top quark pole mass and $m_\tau$.
\item
In addition, NMHDECAY needs the (pole) quark masses $m_s$ and
$m_c$, as well as the CKM matrix elements $V_{us}$, $V_{cb}$ and
$V_{ub}$. The numerical values of these five parameters are 
defined in the subroutine INPUT. (For convenience, they are printed out
in the output file {\tt spectr.dat}, see below.)\par

\een
\noi c) ``BLOCK MINPAR'' contains only the switch 3 with
the input value for $\tan\beta$ specified.\par

\noi d) ``BLOCK EXTPAR'' contains inputs for the \susy\
and soft-\susy-breaking
parameters. Needed generalizations of the SLHA conventions appear.
The new entries are:\newline
\hspace*{2in}{\begin{tabular}{lll}
61 & \hspace*{.5in} & for $\lambda$ \\
62 & \hspace*{.5in} & for $\kappa$ \\
63 & \hspace*{.5in} & for $A_\lambda$ \\
64 & \hspace*{.5in} & for $A_\kappa$
\end{tabular}
}\newline

\noi Note that the entry 23 ($\mu$) is used for the effective $\mu$
parameter, $\mu_\mathrm{eff} = \lambda \left< S \right>$, in the NMSSM.
It should also be noted that neither the 
slepton entries (all generations) nor the squark entries for the first
and second generations are actually used; these are included for future
use only.\par

The two output files of NMHDECAY\_SLHA are {\tt spectr.dat} (see
Table~2) and {\tt
decay.dat}. In  {\tt spectr.dat}, ``BLOCK SPINFO'' is followed by
warnings (switch 3) if any of the constraints described in section 4
are violated, or if any of the six Higgs states could decay into squarks
in which case the corresponding widths and branching ratios should not be
used. This segment of the output also displays error messages
(switch 4) if any of the Higgs, stop or sbottom states have a negative
mass squared. No spectrum output is produced in this case.\par

In the output, ``BLOCK SMINPUTS'' is followed by a printout of the
Standard Model input parameters. The numerical values for $M_W$, $m_s$,
$m_c$, $V_{us}$, $V_{cb}$ and $V_{ub}$, that have no SLHA numbers,
appear in lines subsequent to ``\# SMINPUTS Beyond SLHA''. Similarly,
``BLOCK MINPAR'' is followed by a printout of the value of $\tan\beta$
and ``BLOCK EXTPAR'' is followed by a printout of the important \susy\
and soft-\susy-breaking parameters.  \par

The (pole) masses for the Higgs particles, the neutralinos, the
charginos and squarks appear in the output following ``BLOCK MASS''.
There, one finds several essential NMSSM generalizations of the SLHA
conventions. The new entries, with proposed PDG codes, are \bigskip

\hspace*{1in}
\begin{tabular}{lll}
 45 & \hspace*{1in} & for the third CP-even Higgs boson, \\
 46 & \hspace*{1in}  & for the second CP-odd Higgs boson, \\
 1000045 &\hspace*{1in} & for the fifth neutralino. \\
\end{tabular}
\medskip

\noi The Higgs mixings in the CP-even sector follow ``BLOCK HIGMIX''
 and those in the CP-odd sector follow ``BLOCK AMIX''.
Both segments of the output contain NMSSM generalizations of the SLHA 
conventions, that are required in order to parameterize the mixing in the
enlarged Higgs sector. The meaning of the matrix elements
$S_{ij}$ (i, j = 1, 2, 3) and $P_{ij}'$ (i, j = 1, 2) is as follows.\par
\bit
\item
According to the SLHA conventions the Higgs states $H_u$, $H_d$ are
denoted by $H_2$, $H_1$, respectively. (Inside the Fortran code the
Higgs states $H_u$, $H_d$ are denoted by $H_1$, $H_2$, which is of no
relevance for the SLHA output.)  Hence, for the purpose of the SLHA
output, the CP-even Higgs weak eigenstates are numbered by $S^{weak}_i
= (H_{dR}, H_{uR}, S_R)$ ($R$ refers to the real component of the
field). If  $h_i$ are the mass eigenstates (ordered in mass), the
convention is $h_i = S_{ij} S^{weak}_j$. \par

\item In the CP-odd sector the bare  Higgs fields are $H_{uI}, H_{dI},
S_I$ ($I$ for imaginary component). 
Again, for the purpose of the SLHA output, the CP-odd Higgs weak
eigenstates are denoted by $H_{uI} = H_{2I}$, $H_{dI} = H_{1I}$.
The mass eigenstates are $a_i$ (ordered in mass, $i=1, 2$) and the
Goldstone mode $\tilde G$. Then the elements of $P_{ij}'$ are defined
as 
\bea
H_{1I} &=& \sin\b (P_{11}' a_1 + P_{21}' a_2) 
+ \cos\b \tilde{G} \ ,\nn \\
H_{2I} &=& \cos\b (P_{11}' a_1 + P_{21}' a_2) 
- \sin\b \tilde{G} \ ,\nn \\
S_I &=& P_{12}' a_1 + P_{22}' a_2\ .
\eea
\item
In the output, ``BLOCK NMIX'' is followed by a printout
of the obvious generalization of the $4 \times 4$ MSSM
neutralino mixing matrix to the $5 \times 5$ NMSSM neutralino mixing
matrix (with real entries); ``BLOCK UMIX'' and ``BLOCK VMIX'' 
are followed by printouts of the $U$ and $V$ matrices as defined 
in the MSSM.\par
\item
The output file {\tt decay.dat} respects the SLHA conventions for the Decay
file (at present we consider two particle final states only), using the
above generalizations of the PDG codes both for the decaying particle
and the final states.
\eit

\subsection{NMHDECAY\_SCAN}

NMHDECAY\_SCAN uses the input files {\tt smpar.dat} and 
{\tt   scaninp.dat},  versions 
of which are downloaded automatically with the Fortran code. One
of the sample files appears in Table~\ref{scaninpcase2}.\par

In the file {\tt smpar.dat} the following standard model 
parameters must be specified: 
$\alpha_s(M_Z)$, $G_F$, $\alpha_\mathrm{e.m.}^{-1}(0)$, the lepton
masses $m_\tau$ and $m_\mu$, $M_Z$, $M_W$, the pole quark masses
$m_s$, $m_c$, the running bottom quark mass $m_b(m_b)$, the top quark
pole mass $m_t$, and the CKM matrix elements $V_{us}$, $V_{cb}$ and
$V_{ub}$.\par

In the file {\tt scaninp.dat}, the following must be specified:
\bit
\item the total number of points to be scanned in parameter space;
\item the output format (0 for ``short'', corresponding to simple rows of
numbers per allowed point in parameter space, and 1 for ``long'', as
described below);
\item lower and upper limits for the NMSSM parameters $\lambda$,
$\kappa$, $\tan\beta$, $\mu_\mathrm{eff}$, $A_{\lambda}$ and
$A_{\kappa}$;
\item
 the soft squark masses, trilinear couplings and gaugino masses
over which no scan is performed --- the slepton masses as well as the
squark masses of the first two generations (without index 3) are not
(yet) used.
\eit
The scan in parameter space uses a random number generator, such that
all NMSSM parameters are randomly chosen point by point in the parameter
space within the specified limits.

The output file containing the physical parameters is always called
{\tt scanout.dat}, regardless of  the output format chosen. 
The numbers printed
out for the output format 0 (recommended for scans over more than ~10
points in parameter space) should be edited according to the user's
needs.\footnote{See the section of the program following the
comment line `` The following 3 lines should be edited according to
the user's needs''.}
 The output format 1 is easily readable and shows

\noi - the NMSSM parameters for each point,

\noi - possible warnings in case any phenomenological constraint is
violated, a strong coupling regime below $M_\mathrm{GUT}$ appears or an
unphysical minimum is deeper than the physical one (cf. section 4), if
any of the six Higgs states could decay into squarks (in which case the
corresponding widths and branching ratios should not be used) or error
messages ("fatal" errors) in case any of the Higgs, stop or sbottom
states has a negative mass squared (in which case no additional output
is produced).

\noi - for each of the six Higgs states, their mass, their decomposition
into weak eigenstates $(H_u,\ H_d,\ S)$ (in {\tt Components}), 
their reduced
couplings to gauge bosons (CV), up type quarks (CU), down type quarks
(CD), two gluons (CG) and two photons (CGA) (all relative to a standard
model Higgs boson with the same mass), their branching ratios (where
"Higgses" denote all possible two Higgs final states, and "sparticles"
all possible two particle neutralino/chargino final states), and their
total width,

\noi - the neutralino and chargino masses (all masses in GeV),
as well as the neutralino composition in the basis 
$\psi^0 = (-i\l_1 , -i\l_2, \psi_u^0, \psi_d^0,\psi_s)$
described in detail in Appendix A.

\noi - the (pole) stop and sbottom masses (in GeV).

The output file {\tt scanerr.dat} shows how many of the points in parameter
space have avoided fatal errors or violations of phenomenological
constraints, and the range in the NMSSM parameter space over which points
have passed all these tests.

For users who wish to call a subroutine as a function of the
Higgs, chargino, neutralino and squark sector outputs,
including mixing angles  
and other parameters and quantities computed during the course of the scan,
they should use the parameters and common blocks found in
NMHDECAY\_SCAN.f within ``SUBROUTINE OUTPUT''.  The comments
included in this subroutine should allow easy identification of
all the parameters, branching ratios, mixing angles and so forth
that would be of potential interest for inputting into 
a user's subroutine.

\mysection{Results and discussion}
\label{sec:results}

As stated above, the masses, couplings and decay properties of the Higgs
bosons of the NMSSM can differ significantly from the MSSM. The primary
purpose of NMH\-DECAY is to allow for detailed studies of such cases.\par

A particularly interesting case is where the lightest Higgs scalar $h_1$,
although primarily non-singlet, decays mainly into two pseudoscalars
(also primarily non-singlet) \cite{higsec2}. It is then interesting
to see how the branching ratios of $h_1$ depend on the parameters of the
model. In particular, we can determine whether the choices of parameters 
that yield the above types of decays are ruled out for some of the
phenomenological reasons discussed in section 4. We can also
determine if there are particular
(fine-tuned) relations between the parameters required for $h_1\to
a_1a_1$ to be dominant. To study such issues, it will be convenient
to fix all but one of the
parameters (which allows for a reasonable graphical representation), and
perform a scan over the remaining parameter. \par

In some sense, the input parameter $A_{\lambda}$ is the most natural one to
vary, since the mass of the MSSM like pseudoscalar depends quite strongly
on $A_{\lambda}$ (and hence $A_{\lambda}$ plays the role of $M_A$ in the
MSSM).\par 

\begin{figure}[p]
\begin{center}
\includegraphics[width=4.2in]{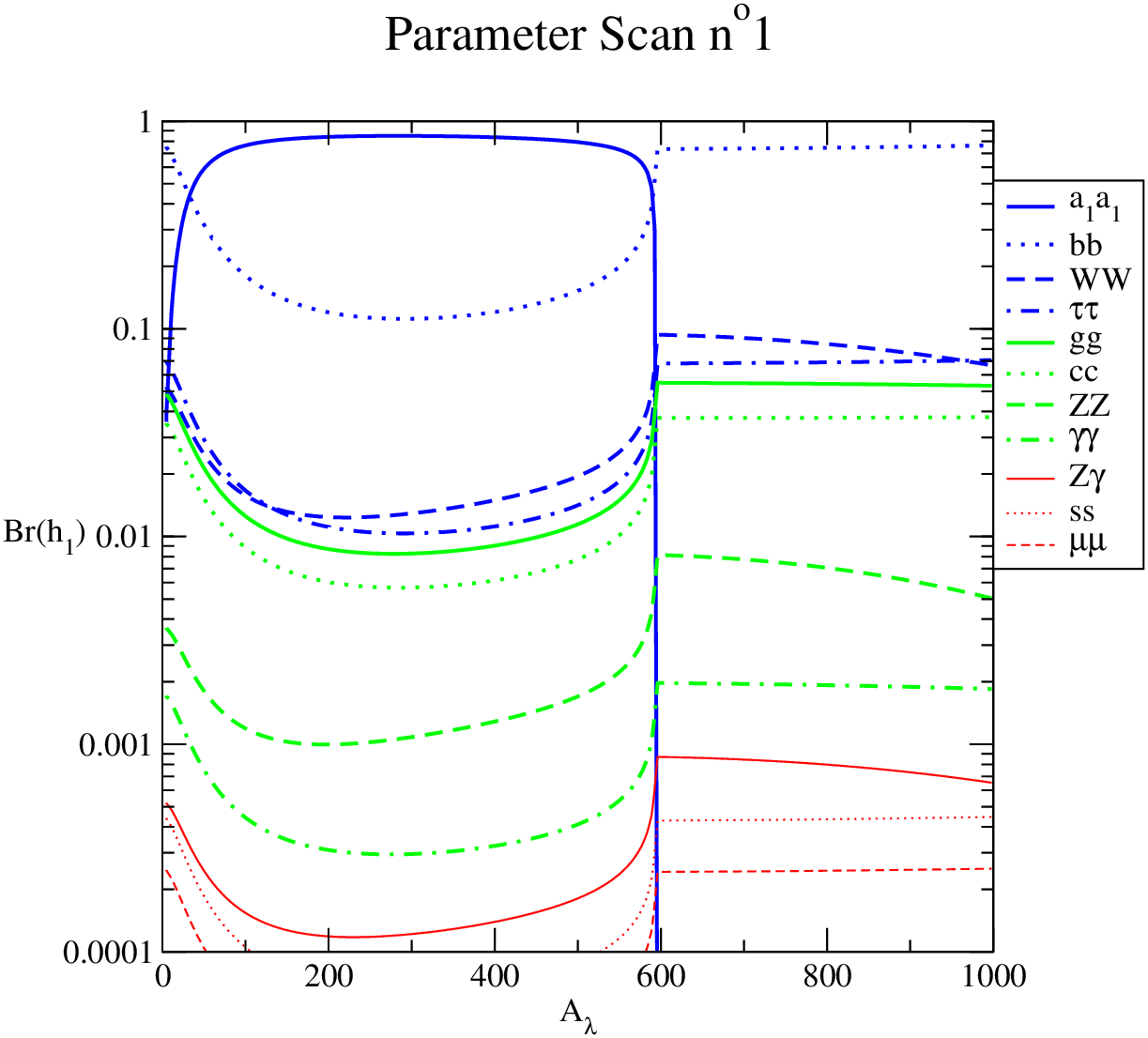}
\caption{Branching ratios of $h_1$ as a function of $A_{\lambda}$ for
$\lambda=\kappa= 0.3$, $\tan\beta=5$, $\mu_\mathrm{eff} = 180\gev$,
$A_{\kappa}$ = 0, $m_\mathrm{squark} =1\tev$, and $A_t=1.5\tev$. 
\label{nmhfig1}}
\vskip.3in

\includegraphics[width=4.2in]{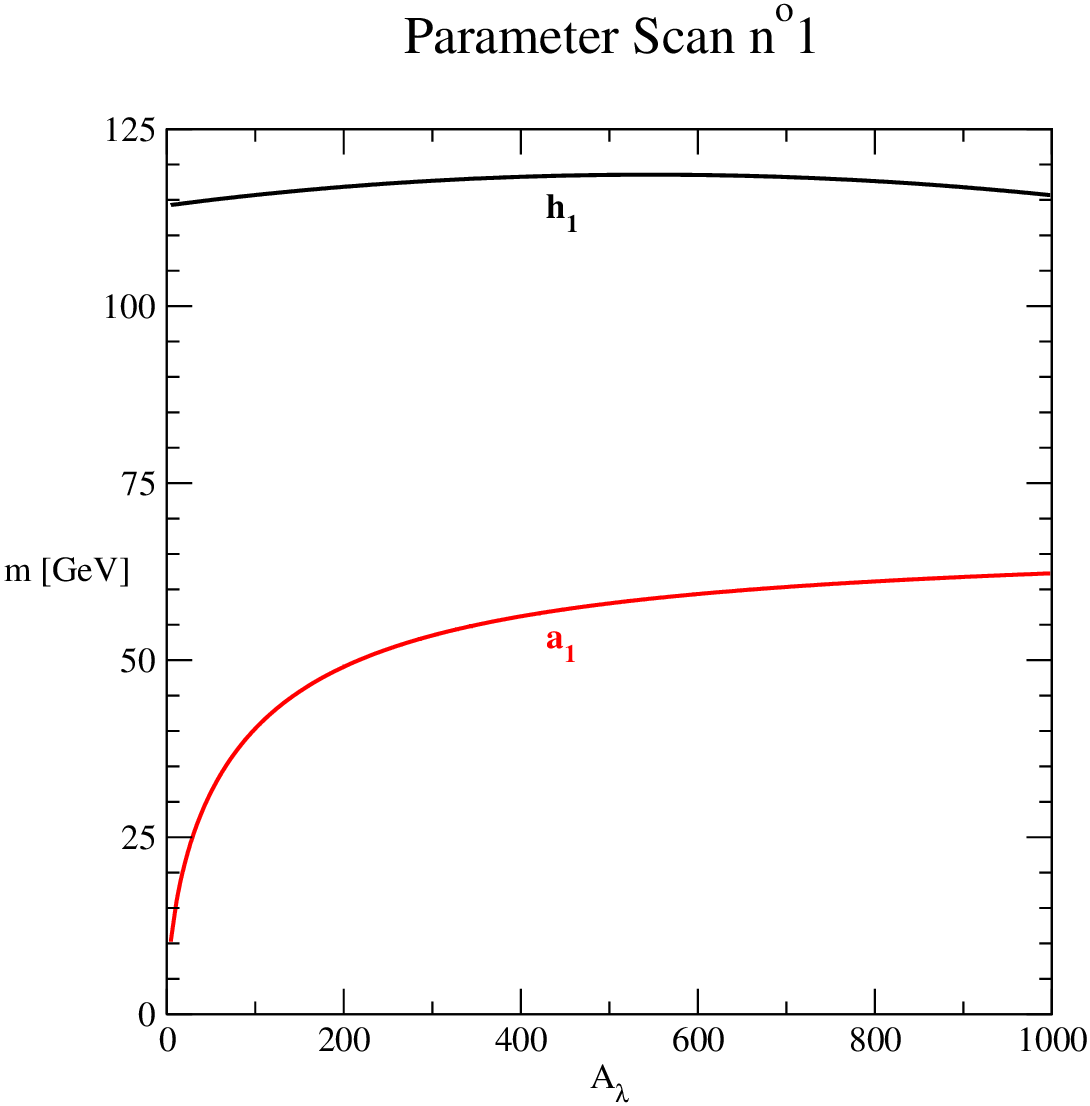}
\end{center}
\vspace*{-.2in}
\caption{$m_{h_1}$ and $m_{a_1}$ as a function of $A_{\lambda}$ for
the same parameters as in fig.~\ref{nmhfig1}.\label{nmhfig2}}
\end{figure}

Let us first consider the following choice of the NMSSM parameters
[cf. eq.~(\ref{6param})]: $\lambda = \kappa= 0.3$, $\tan\beta$ = 5, $\mu_\mathrm{eff}$ = 180~GeV,
$A_{\kappa}$ = 0. For the squark masses and trilinear couplings, we take 
1~TeV and $1.5\tev$, respectively. 
Varying $A_{\lambda}$ between 0 and 1000~GeV, we obtain the
branching ratios for $h_1$ as shown in fig.~\ref{nmhfig1}. 
These show clearly that,
for $A_{\lambda}\ \lsim \ 600$~GeV, the decay $h_1 \to a_1\ a_1$ is
dominant.

The reason for the sharp drop of the $h_1\to a_1a_1$
branching ratio for
$A_{\lambda}\ > \ 600$~GeV becomes clear from fig.~\ref{nmhfig2},
where we show the
masses $m_{h_1}$ and $m_{a_1}$ as functions of $A_{\lambda}$; for
$A_{\lambda}\ > \ 600$~GeV, $m_{a_1}$ becomes larger than $m_{h_1}/2$.
The $h_1\to a_1a_1$ decay is also reduced as $A_\lambda\to 0$,
even though the $a_1$ 
becomes very light, because in this case it is mainly singlet and has a very
small coupling to the $h_1$. 
None of the points in these two graphs are excluded by LEP.
They should be visible at the LHC using the techniques we have developed
\cite{higsec2}
for isolating the $WW\to h \to aa$ type of signal.

\begin{figure}[p]
\begin{center}
\includegraphics[width=4.5in]{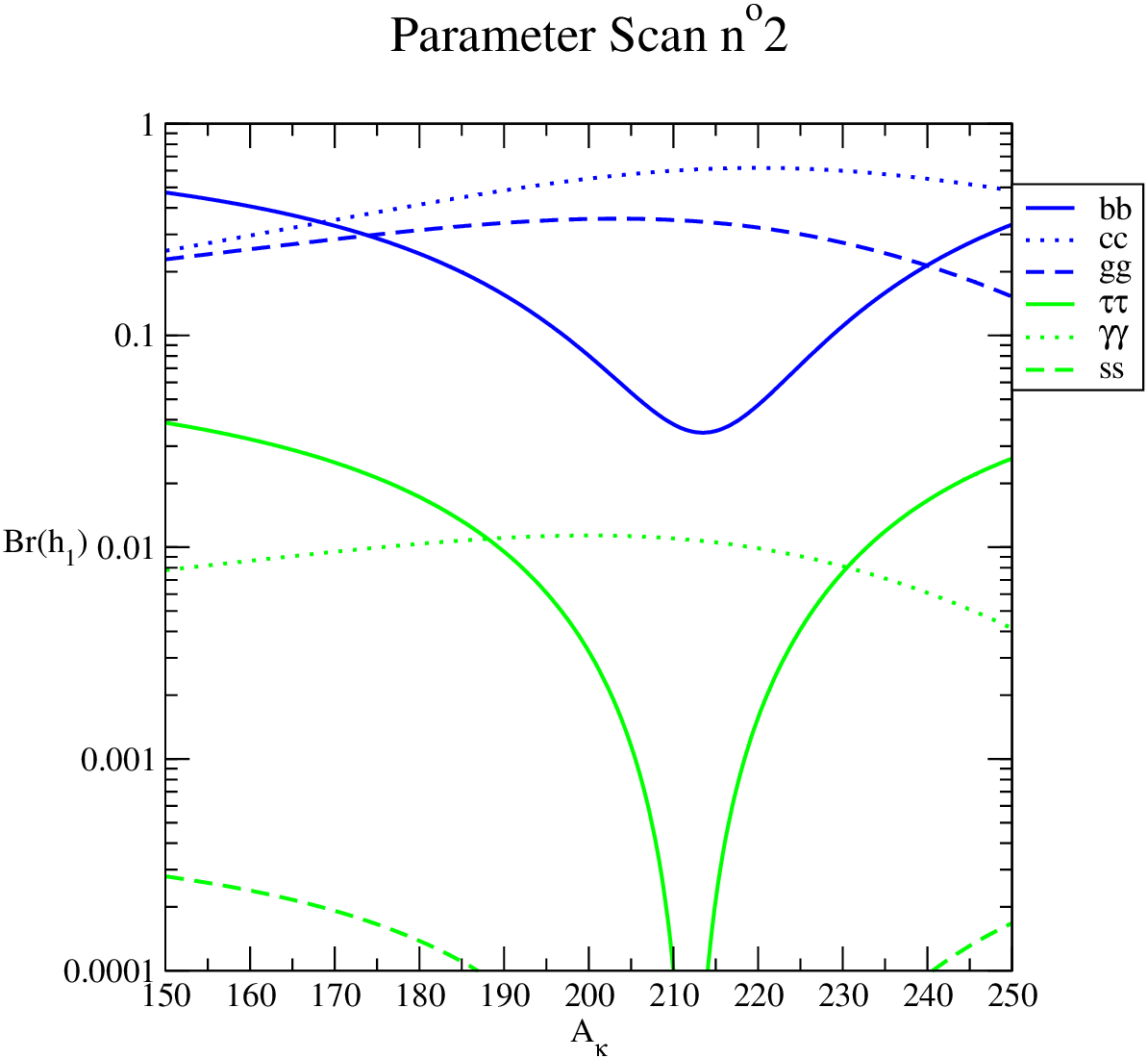}
\caption{Branching ratios of $h_1$ as a function of $A_{\kappa}$ for
$\lambda=0.5$, $\kappa=-0.15$, $\tan\beta$ = 3.5, 
$\mu_\mathrm{eff}$ = 200~GeV, $A_{\lambda}=780\gev$, 
$m_\mathrm{squark} =1\tev$, and $A_t=1.5\tev$. \label{nmhfig3}}
\vskip.27in

\includegraphics[width=4.5in]{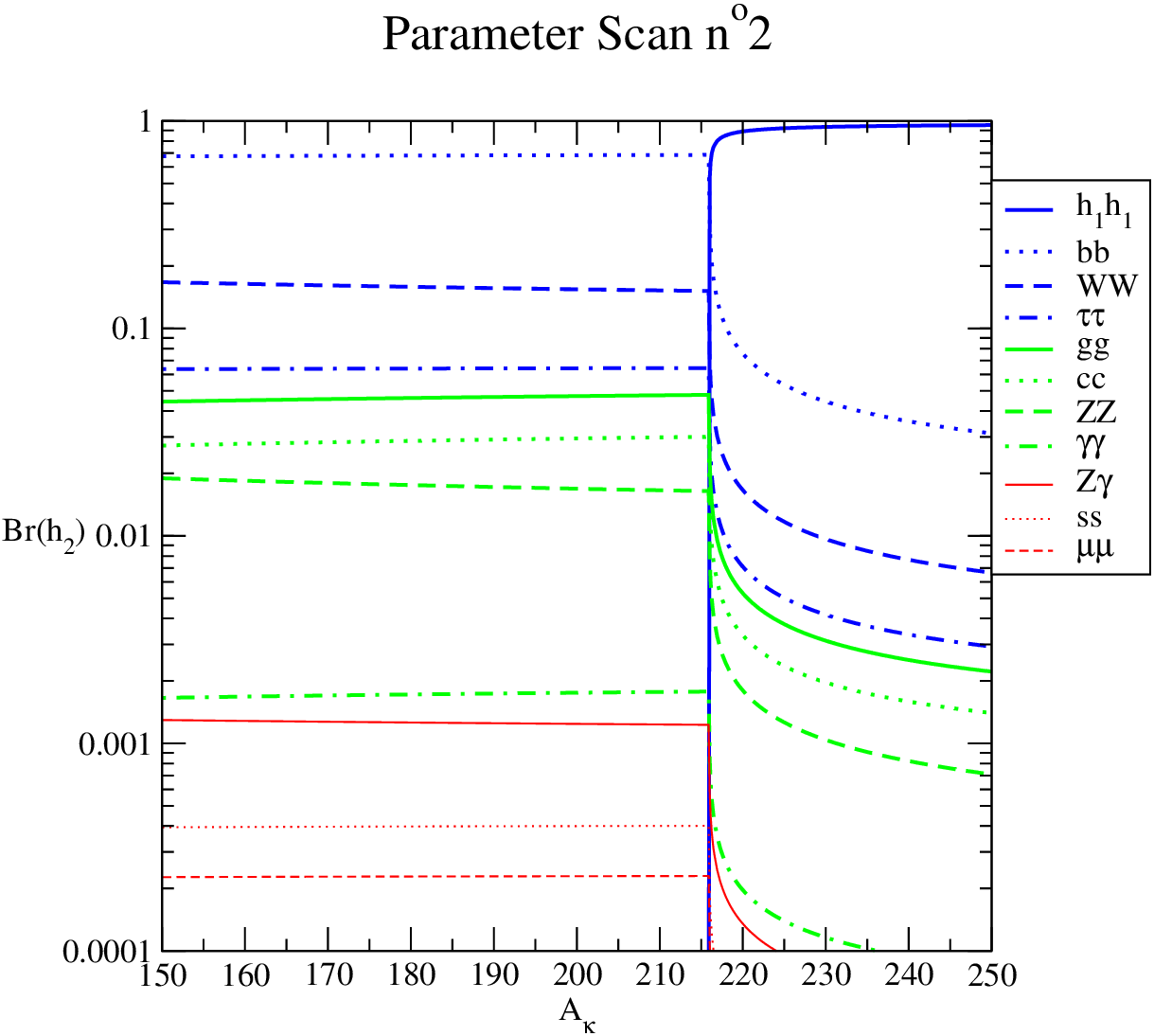}
\end{center}
\vspace*{-.2in}
\caption{\label{nmhfig4}Branching ratios of $h_2$ as a function of
$A_{\kappa}$ for the same parameter choices as in fig.~\ref{nmhfig3}.}
\end{figure}

\begin{figure}[p]
\begin{center}
\includegraphics[width=4.in,height=4.2in]{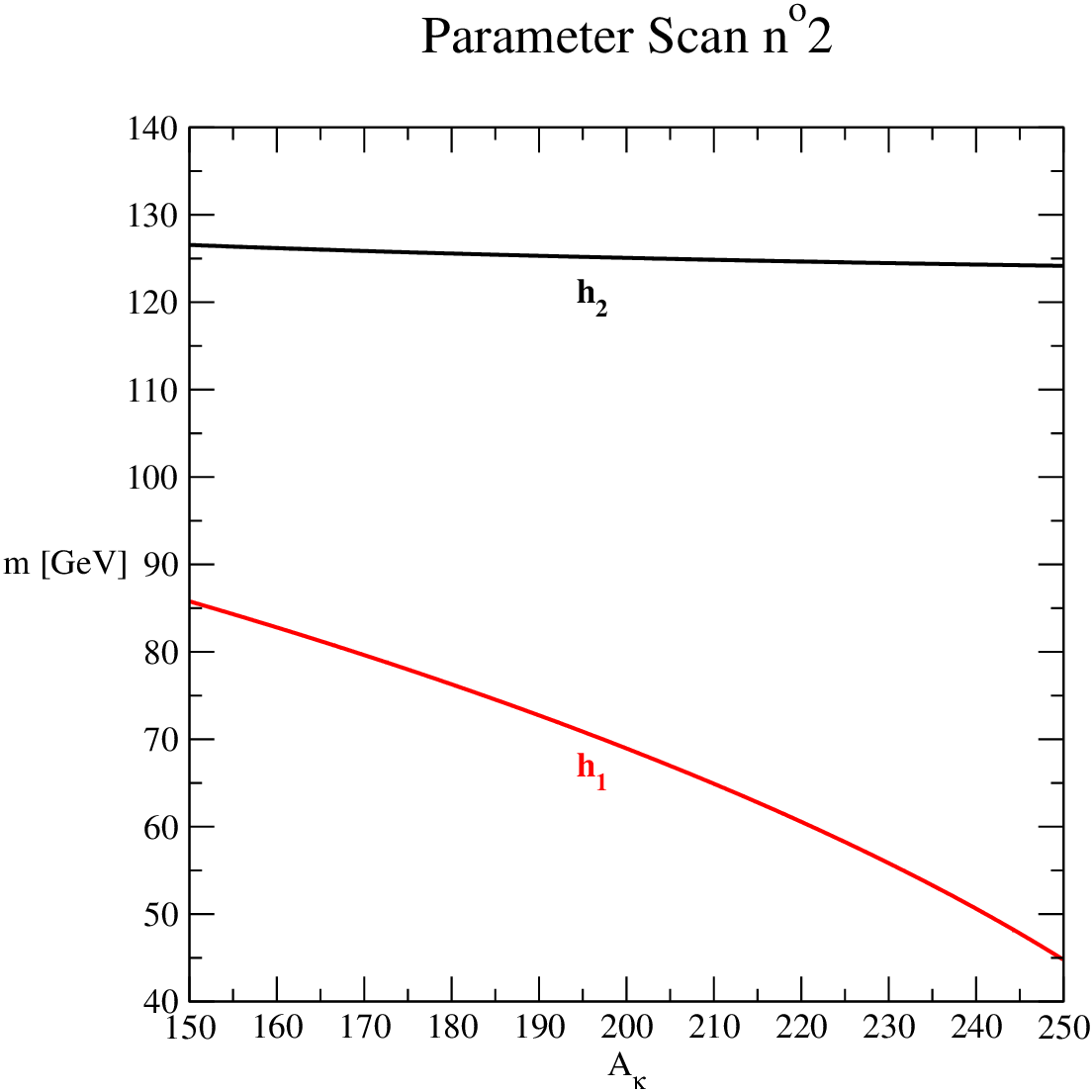}
\end{center}
\vspace*{-.25in}
\caption{$m_{h_1}$ and $m_{h_1}$ as a function of $A_{\kappa}$ for
the same parameters as in fig.~\ref{nmhfig3}.\label{nmhfig5}}
\vskip.3in
\begin{center}

\includegraphics[width=3.7in,height=4.2in]{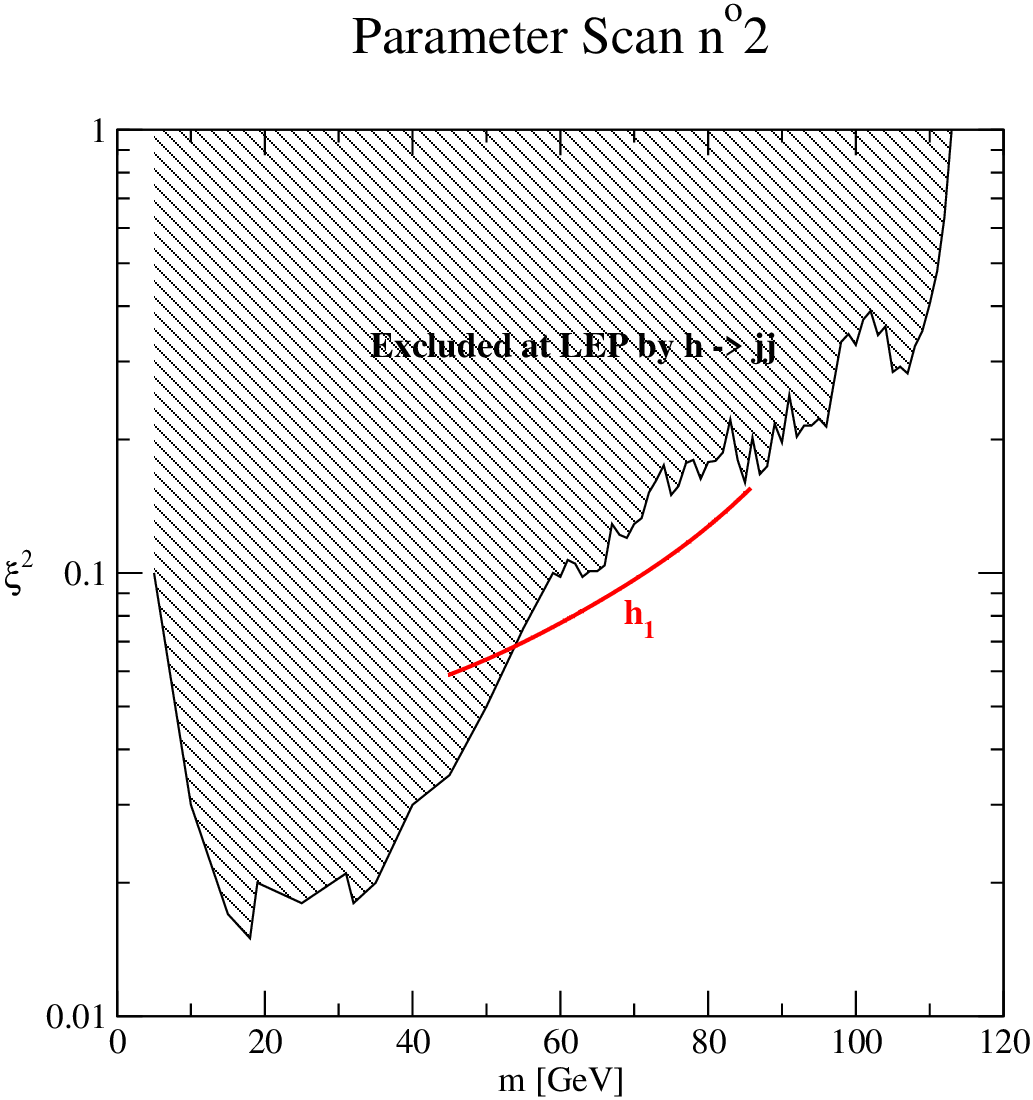}
\end{center}
\vspace*{-.2in}
\caption{LEP constraints in comparison to predictions for $h_1$
for the parameters of fig.~\ref{nmhfig3}. Note the correlation
of $m=m_{h_1}$ with $A_\kappa$ given in
fig.~\ref{nmhfig5}. \label{nmhfig6}}
\end{figure}

For a second sample set of plots,
figs.~\ref{nmhfig3}--\ref{nmhfig6}, we take $\lambda=0.5$, 
$\kappa=-0.15$, $\tan\beta=3.5$, 
$\mu_\mathrm{eff}=200\gev$, $A_\lambda=780\gev$ and $A_\kappa\in [150\gev,
250\gev]$. The {\tt scaninp.dat} file for this
case is given in Table~\ref{scaninpcase2}.
For much of this parameter range, neither the $h_1$
nor the $h_2$ would have been observable at LEP. In particular,
fig.~\ref{nmhfig5} shows that 
$m_{h_2}\ \gsim\ 120\gev$ implying that the $h_2$ is beyond
the LEP kinematical reach. The $h_1$ is much lighter.
However, this light
Higgs is not excluded by LEP over most of the above $A_\kappa$ range
since: a) its reduced coupling to gauge
bosons is small; and b) $h_1\to b\overline b$ is
suppressed so that $h_1\to jj$ decays are dominant (see
fig.~\ref{nmhfig3}). In fig.~\ref{nmhfig6},  
we plot $\xi^2=C_V(h_1)^2 \times BR(h_1\to jj)$ for our
selected points as well as the region excluded by LEP searches
in this channel \cite{LHWGjj}. We see that
only if $m_{h_1}\ \lsim\ 53\gev$, which corresponds to $A_\kappa\ \gsim\
235\gev$, would the $h_1$ be excluded by LEP data. 

Will these Higgs bosons be observable at the LHC? In this regard,
it is important to note from fig.~\ref{nmhfig4} that
when $A_\kappa\ \gsim\ 215\gev$, $h_2\to h_1h_1$ decays are dominant.
This occurs because $m_{h_1}$ decreases
with $A_\kappa$, see fig.~\ref{nmhfig5}.
Meanwhile, fig.~\ref{nmhfig3} shows that
$BR(h_1\to b\bar b)$ and $BR(h_1\to\tau^+\tau^-)$ are both small when
$A_\kappa\in[205\gev,220\gev]$; in this region of parameter space,
the $h_1$ decays mainly to $c\bar c$ or $gg$. 
Thus, for $A_\kappa\sim 215-220\gev$:
\bit
\item The $h_1$ has a mass that lies below the mass range currently
studied for Higgs detection at the LHC. Further, the
$h_1$ will be so weakly produced
at the LHC (since $\xi^2\ \lsim\ 0.1$) that extensions to lower Higgs masses
of the current LHC studies would probably conclude it was undetectable.
\item
Simultaneously, 
the strongly produced $h_2$ has decays dominated by $h_2\to h_1 h_1$
with $h_1\to c\bar c,gg$ (but not $b\bar b$ or $\tau^+\tau^-$).
As a result, the techniques of
\cite{higsec2} for $h\to aa$ (which
require a significant $a\to \tau^+\tau^-$ branching ratio)
do not apply, and the $h_2$ would also appear to be very difficult to
observe at the LHC. 
\eit
We note that the above choice of parameters
produces a phenomenology for the Higgses that is somewhat similar
to that discussed in \cite{higsec3}, which focuses on
a region of parameter space such that the $h_1$ has suppressed
decays to $b\bar b$ and $\tau^+\tau^-$.  The difference is
that the decay of their $h_2$ to $h_1h_1$ is kinematically forbidden
and the $h_2$ could be detected at the LHC. 

\mysection{Conclusions}
\label{sec:conclusions}

In this paper, we have presented details regarding the now
publicly available \break
NMHDECAY programs, NMHDECAY\_SLHA.f and
NMHDECAY\_SCAN.f, that can be used to explore the Higgs
sector of the Next to Minimal Supersymmetric Model defined
by adding one singlet Higgs superfield to the Minimal
Supersymmetric Model. The programs, and associated data files, can
be downloaded from the two web pages:

\hspace*{.5in}{\tt{http://www.th.u-psud.fr/NMHDECAY/nmhdecay.html}}

\hspace*{.5in}{\tt{http://higgs.ucdavis.edu/nmhdecay/nmhdecay.html}}

\noi The web pages provide simplified descriptions
of the programs and instructions on how to use them.
The programs will be updated to include additional
features and refinements in subsequent versions.  We welcome comments
with regard to improvements that users would find helpful.

We have exemplified the use of these programs for two particularly
interesting scenarios: The first illustrates
the potentially crucial importance of the LHC $h\to aa$ detection mode
(that is dominant over a significant, not fine-tuned, range of
parameters of the NMSSM).  The second exposes a limited portion of
parameter space for which Higgs discovery would not 
have occurred at LEP and will
probably not be possible at the LHC.

\section*{Acknowledgments}

We thank P. Skands for comments on new SLHA and PDG particle codes, and
M. Boonekamp, K. Desch, P. Ferrari, B. Murray, M. Oreglia and A. L. Read
for communications on experimental results at LEP.
JFG is supported by the U.S. Department of Energy.
CH is supported by the European Commission RTN grant
HPRN-CT-2000-00148.
The support for this work received 
from the France Berkeley Fund
was critical from inception to completion.

\begin{table}[p]
{\footnotesize
\begin{verbatim}
#  INPUT FILE FOR NMHDECAY 
#  BASED ON SUSY LES HOUCHES ACCORD, MODIFIED FOR THE NMSSM
#  IN EXTPAR: LINES 61-64: NMSSM YUKAWA COUPLINGS AND TRILIN. SOFT TERMS
BLOCK MODSEL
  3     1     # NMSSM PARTICLE CONTENT
BLOCK SMINPUTS
  1   137.036 # ALPHA_EM^-1(0)
  2   1.16639D-5   # GF
  3   0.12    # ALPHA_S(MZ)
  4   91.187  # MZ
  5     4.24  # MB(MB), RUNNING B QUARK MASS
  6   175.    # TOP QUARK POLE MASS
  7   1.7771  # MTAU
BLOCK MINPAR
  3    5.     # TANBETA
BLOCK EXTPAR
  1   5.D2    # M1
  2   1.D3    # M2
  3   3.D3    # M3
  11  1.5D3   # ATOP
  12  1.5D3   # ABOT
  13  1.5D3   # ATAU
  23  180.    # MU
  31  1.D3    # LEFT SELECTRON
  32  1.D3    # LEFT SMUON
  33  1.D3    # LEFT STAU
  34  1.D3    # RIGHT SELECTRON 
  35  1.D3    # RIGHT SMUON
  36  1.D3    # RIGHT STAU
  41  1.D3    # LEFT 1ST GEN. SQUARKS
  42  1.D3    # LEFT 2ND GEN. SQUARKS
  43  1.D3    # LEFT 3RD GEN. SQUARKS
  44  1.D3    # RIGHT U-SQUARKS
  45  1.D3    # RIGHT C-SQUARKS
  46  1.D3    # RIGHT T-SQUARKS
  47  1.D3    # RIGHT D-SQUARKS
  48  1.D3    # RIGHT S-SQUARKS
  49  1.D3    # RIGHT B-SQUARKS
  61  .3D0    # LAMBDA
  62  .3D0    # KAPPA
  63  200.    # A_LAMBDA
  64  0.0     # A_KAPPA
\end{verbatim}}
\caption{Sample {\tt slhainp.dat} file.\label{slhainp}}
\end{table}

\begin{table}
{\footnotesize
\baselineskip 10pt
\begin{verbatim}
# NMHDECAY OUTPUT IN SLHA FORMAT
# Info about spectrum calculator
BLOCK SPINFO        # Program information
     1   NMHDECAY   # spectrum calculator
     2   1.1        # version number
# Input parameters
BLOCK MODSEL
    3     1         # NMSSM PARTICLE CONTENT
BLOCK SMINPUTS
     1     1.37036000E+02   # ALPHA_EM^-1
     2     1.16639000E-05   # GF
     3     1.20000000E-01   # ALPHA_S(MZ)
     4     9.11870000E+01   # MZ
     5     4.24000000E+00   # MB(MB)
     6     1.75000000E+02   # MTOP (POLE MASS)
     7     1.77710000E+00   # MTAU
# SMINPUTS Beyond SLHA:
# MW:     0.80419998E+02
# MS:     0.19000000E+00
# MC:     0.16100000E+01
# VUS:     0.22200000E+00
# VCB:     0.41000000E-01
# VUB:     0.36000000E-02
BLOCK MINPAR
     3     5.00000000E+00   # TANBETA
BLOCK EXTPAR
     1     5.00000000E+02   # M1
     2     1.00000000E+03   # M2
     3     3.00000000E+03   # M3
    11     1.50000000E+03   # ATOP
    12     1.50000000E+03   # ABOTTOM
    13     1.50000000E+03   # ATAU
    23     1.80000000E+02   # MU
    33     1.00000000E+03   # LEFT STAU
    36     1.00000000E+03   # RIGHT STAU
    43     1.00000000E+03   # LEFT 3RD GEN. SQUARKS
    46     1.00000000E+03   # RIGHT T-SQUARKS
    49     1.00000000E+03   # RIGHT B-SQUARKS
    61     3.00000000E-01   # LAMBDA
    62     3.00000000E-01   # KAPPA
    63     2.00000000E+02   # A_LAMBDA
    64     0.00000000E+00   # A_KAPPA
# 
BLOCK MASS   # Mass spectrum 
#  PDG Code     mass             particle 
        25     1.16801961E+02   # lightest neutral scalar
        35     3.56992533E+02   # second neutral scalar
        45     5.91271441E+02   # third neutral scalar
        36     4.90592823E+01   # lightest pseudoscalar
        46     5.87736637E+02   # second pseudoscalar
        37     5.90714771E+02   # charged Higgs
   1000022     1.66456703E+02   # neutralino(1)
   1000023    -1.85971845E+02   # neutralino(2)
   1000025     3.67762262E+02   # neutralino(3)
   1000035     5.04634098E+02   # neutralino(4)
   1000045     1.00711878E+03   # neutralino(5)
   1000024     1.76259627E+02   # chargino(1)
   1000037     1.00710838E+03   # chargino(2)
   1000005     8.21132887E+02   #  ~b_1
   1000006     6.24445725E+02   #  ~t_1
   2000005     8.24131926E+02   #  ~b_2
   2000006     1.00488215E+03   #  ~t_2
\end{verbatim}}
\end{table}

\begin{table}[p]
{\footnotesize
\baselineskip 10pt
\begin{verbatim}
# 3*3 Higgs mixing
BLOCK HIGMIX
  1  1    -1.27437514E-01   # S_(1,1)
  1  2    -3.16386555E-02   # S_(1,2)
  1  3    -9.91341856E-01   # S_(1,3)
  2  1    -2.01793548E-01   # S_(2,1)
  2  2    -9.77759539E-01   # S_(2,2)
  2  3     5.71458423E-02   # S_(2,3)
  3  1    -9.71101974E-01   # S_(3,1)
  3  2     2.07328914E-01   # S_(3,2)
  3  3     1.18218769E-01   # S_(3,3)
# 2*2 Pseudoscalar Higgs mixing
BLOCK AMIX
  1  1    -2.37325986E-02   # P'_(1,1)
  1  2    -9.99718342E-01   # P'_(1,2)
  2  1     9.99718342E-01   # P'_(2,1)
  2  2    -2.37325986E-02   # P'_(2,2)
# Gaugino-Higgsino mixing
BLOCK NMIX  # 5*5 Neutralino Mixing Matrix
  1  1     1.05202671E-01   # N_(1,1)
  1  2    -7.87589631E-02   # N_(1,2)
  1  3     6.92565025E-01   # N_(1,3)
  1  4    -6.93980327E-01   # N_(1,4)
  1  5     1.46541512E-01   # N_(1,5)
  2  1    -3.48143940E-02   # N_(2,1)
  2  2     3.76738427E-02   # N_(2,2)
  2  3     7.01102969E-01   # N_(2,3)
  2  4     7.06807991E-01   # N_(2,4)
  2  5     7.90299289E-02   # N_(2,5)
  3  1    -2.45218320E-02   # N_(3,1)
  3  2     9.59570918E-03   # N_(3,2)
  3  3    -1.58479313E-01   # N_(3,3)
  3  4     4.52366830E-02   # N_(3,4)
  3  5     9.85973910E-01   # N_(3,5)
  4  1     9.93510538E-01   # N_(4,1)
  4  2     1.73884449E-02   # N_(4,2)
  4  3    -5.24484443E-02   # N_(4,3)
  4  4     9.87396900E-02   # N_(4,4)
  4  5     1.15796113E-02   # N_(4,5)
  5  1    -7.47306864E-03   # N_(5,1)
  5  2     9.95983570E-01   # N_(5,2)
  5  3     3.06884365E-02   # N_(5,3)
  5  4    -8.37728195E-02   # N_(5,4)
  5  5    -1.10280322E-03   # N_(5,5)
BLOCK UMIX  # Chargino U Mixing Matrix
  1  1     4.31428408E-02   # U_(1,1)
  1  2    -9.99068914E-01   # U_(1,2)
  2  1     9.99068914E-01   # U_(2,1)
  2  2     4.31428408E-02   # U_(2,2)
BLOCK VMIX  # Chargino V Mixing Matrix
  1  1     1.18343118E-01   # V_(1,1)
  1  2    -9.92972762E-01   # V_(1,2)
  2  1     9.92972762E-01   # V_(2,1)
  2  2     1.18343118E-01   # V_(2,2)
\end{verbatim}}
\caption{Corresponding {\tt spectr.dat} output file.}
\end{table}

\begin{table}[p]
{\footnotesize
\baselineskip 10pt
\begin{verbatim}
#
#  Total number of points scanned
#
1000
#
#  Output format 0=short 1=long (not recommended for big scannings)
#
0
#
#  lambda
#
0.5
0.5
#
#  kappa
#
-0.15
-0.15
#
#  tan(beta)
#
3.5
3.5
#
#  mu
#
200.
200.
#
#  A_lambda
#
780.
780.
#
#  A_kappa
#
150.0
250.0
#
#  Remaining soft terms (no scan)
#
mQ3=  1.D3
mU3=  1.D3
mD3=  1.D3
mL3=  1.D3
mE3=  1.D3
AU3=  1.5D3
AD3=  1.5D3
AE3=  1.5D3
mQ=   1.D3
mU=   1.D3
mD=   1.D3
mL=   1.D3
mE=   1.D3
M1=   5.D2
M2=   1.D3
M3=   3.D3
\end{verbatim}}
\caption{The {\tt scaninp.dat} file for parameter scan \#2.
\label{scaninpcase2}}
\end{table}

\newpage
\renewcommand{\theequation}{A.\arabic{equation}}
\setcounter{equation}{0}
\setcounter{section}{0}
\section*{Appendix A}

\section{General Conventions for the NMSSM}
Below, we define our conventions for the tree level Lagrangian 
of the NMSSM. The superpotential for the Higgs fields, the quarks and
the leptons of the 3rd generation is

\beq
W = h_t \widehat{Q}\cdot \widehat{H}_u \widehat{T}_R^c - h_b
\widehat{Q} \cdot  \widehat{H}_d  \widehat{B}_R^c - h_{\tau}
\widehat{L} \cdot \widehat{H}_d \widehat{L}_R^c +\l \widehat{S}
\widehat{H}_u \cdot \widehat{H}_d + \frac{1}{3} \k \widehat{S}^3 \ .
\eeq 

\noi Hereafter, hatted capital letters denote superfields,
and unhatted capital letters the corresponding (complex) scalar
components. The $SU(2)$ doublets are

\beq
\widehat{Q} = \left(\ba{c} \widehat{T}_L \\ \widehat{B}_L
\ea\right) , \
\widehat{L} = \left(\ba{c} \widehat{\nu}_{\tau L} \\ \widehat{\tau}_L
\ea\right) , \
\widehat{H}_u = \left(\ba{c} \widehat{H}_u^+ \\ \widehat{H}_u^0
\ea\right) , \
\widehat{H}_d = \left(\ba{c} \widehat{H}_d^0 \\ \widehat{H}_d^-
\ea\right) .
\eeq

\noi Products of two $SU(2)$ doublets are defined as, e.g.,

\beq
\widehat{H}_u \cdot \widehat{H}_d = \widehat{H}_u^+ \widehat{H}_d^- 
- \widehat{H}_u^0 \widehat{H}_d^0\ .
\eeq

\noi For the soft \susy\ breaking terms we take

\bea
-{\cal L}_\mathrm{soft} &=
m_\mathrm{H_u}^2 | H_u |^2 + m_\mathrm{H_d}^2 | H_d |^2 + m_\mathrm{S}^2 | S |^2
+m_Q^2|Q^2| + m_T^2|T_R^2| \nn \\
&+m_B^2|B_R^2| +m_L^2|L^2| +m_\mathrm{\tau}^2|L_R^2|
\nn \\
&+ (h_t A_t\ Q \cdot H_u T_R^c - h_b A_b\ Q \cdot H_d B_R^c - h_{\tau}
A_{\tau}\ L \cdot H_d L_R^c\nn \\
& +\l A_\l\ H_u \cdot H_d S + \third \k A_\kappa\ S^3 + \mathrm{h.c.}
)\,. \eea

\section{Higgs Sector at Tree Level}

For completeness, we list here the Higgs potential, tree level Higgs
masses and our conventions for the mixing angles. The tree level Higgs
potential is given by

\bea
V & = & \l^2 (|H_u|^2|S|^2 + |H_d|^2|S|^2 + |H_u \cdot  H_d|^2) +
\k^2|S^2|^2 \nn \\
&& + \l\k (H_u \cdot H_dS^{*2} + \mathrm{h.c.}) 
+ \quart g^2 (|H_u|^2 - |H_d|^2)^2 \nn \\
& & + \half g_2^2 |H_u^+
H_d^{0*} + H_u^0 H_d^{-*}|^2
 + m_\mathrm{H_u}^2|H_u|^2 + m_\mathrm{H_d}^2|H_d|^2 + m_S^2|S|^2 \nn \\
& &+ (\l A_\l H_u \cdot
H_d S + \third \k A_\k\ S^3 + \mathrm{h.c.}) 
\label{vform}
\eea

\noi where 
\beq
g^2 = \half (g_1^2 + g_2^2)\ .
\eeq

\noi Assuming vevs $h_u$, $h_d$ and $s$ such that
\beq
H_u^0 = h_u + \frac{H_{uR} + iH_{uI}}{\sqrt{2}} , \q
H_d^0 = h_d + \frac{H_{dR} + iH_{dI}}{\sqrt{2}} , \q
S = s + \frac{S_R + iS_I}{\sqrt{2}}
\eeq

\noi eq.~(\ref{vform}) simplifies to
\bea
V & = & \l^2 (h_u^2 s^2 + h_d^2 s^2 + h_u^2 h_d^2) + \kappa^2 s^4
- 2 \l\k h_u h_d s^2 - 2\lambda A_\l\ h_u h_d s \nn\\
& & + \frac{2}{3}\,\kappa A_\k  s^3
+ m_\mathrm{H_u}^2 h_u^2 + m_\mathrm{H_d}^2 h_d^2 + m_S^2 s^2
+ \quart g^2 (h_u^2 - h_d^2)^2\ .
\eea

The sign conventions for the fields 
can be chosen such that the Yukawa couplings $\lambda$, $h_t$,
$h_b$, the vevs $h_u$, $h_d$ (and hence $\tan\beta$) as well as the soft
gaugino masses $M_i$ are all positive. Then, the Yukawa coupling $\kappa$,
the trilinear soft terms $A_i$, and the vev $s$ (and hence $\mu_{eff}$)
can all be either positive or negative.

\subsection{CP-even neutral states}

In the basis $S^{bare} = (H_{uR}, H_{dR}, S_R)$ and using the
minimization equations in order to eliminate the soft masses
squared, one obtains the following mass-squared matrix entries:
\bea
{\cal M}_{S,11}^2 & = & g^2 h_u^2 + \l s \frac{h_d}{h_u}\, (A_\l + \k
s), \nn\\
{\cal M}_{S,22}^2 & = & g^2 h_d^2 + \l s \frac{h_u}{h_d}\, (A_\l + \k
s), \nn\\
{\cal M}_{S,33}^2 & = & \l A_\l \frac{h_u h_d}{s}\, + \k s (A_\k + 4 \k
s), \nn\\
{\cal M}_{S,12}^2 & = & (2\l^2 - g^2) h_u h_d - \l s (A_\l + \k s),
\nn\\ {\cal M}_{S,13}^2 & = & 2\l^2 h_u s - \l h_d (A_\l + 2\k s),
\nn\\
{\cal M}_{S,23}^2 & = & 2\l^2 h_d s - \l h_u (A_\l + 2\k s).
\eea

\noi  After diagonalization by an orthogonal matrix $S_{ij}$ one
obtains 3 CP-even states (ordered in mass) $h_i = S_{ij} S^{bare}_j$,
with  masses denoted by $m_{h_i}$. \par
In the MSSM limit ($\l$, $\k \to 0$, and parameters such that $h_3 \sim
S_R$) the elements of the first $2 \times 2$ sub-matrix of $S_{ij}$ are
related to the MSSM angle $\alpha$ as

\bea
S_{11} \sim &\cos\alpha\ , \qquad &S_{21} \sim \sin\alpha\ ,\nn \\
S_{12} \sim &-\sin\alpha\ , \qquad &S_{22} \sim \cos\alpha\ .
\eea

\subsection{CP-odd neutral states}

In the basis $P^{bare} = (H_{uI}, H_{dI}, S_I)$ and using the
minimization equations in order to eliminate the soft masses
squared, one obtains the following mass-squared matrix entries:
\bea
{\cal M}_{P,11}^2 & = & \l s \frac{h_d}{h_u}\, (A_\l + \k s), \nn\\
{\cal M}_{P,22}^2 & = & \l s \frac{h_u}{h_d}\, (A_\l + \k s), \nn\\
{\cal M}_{P,33}^2 & = & 4 \l \k h_u h_d + \l A_\l \frac{h_u h_d}{s}\,
-3 \k A_\k s, \nn\\
{\cal M}_{P,12}^2 & = & \l s (A_\l + \k s), \nn\\
{\cal M}_{P,13}^2 & = & \l h_d (A_\l - 2\k s), \nn\\
{\cal M}_{P,23}^2 & = & \l h_u (A_\l - 2\k s).
\eea

\noi The diagonalization of this mass matrix is performed in two steps.
First, one rotates into a basis ($\tilde{A}, \tilde{G}, S_I$), 
where $\tilde{G}$ is a massless Goldstone mode:

\beq
\left(\ba{c}H_{uI} \\  H_{dI} \\ S_I \ea\right) = 
 \left(\ba{ccc} \cos\b & -\sin\b & 0 \\ 
 \sin\b & \cos\b & 0 \\
 0 & 0 & 1 \ea\right)
\left(\ba{c} \tilde{A} \\ \tilde{G} \\  S_I \ea\right)
\eeq

\noi where $\tan\beta = h_u/h_d$. Dropping the Goldstone mode, the
remaining $2 \times 2$ mass matrix in the basis ($\tilde{A}, S_I$)
has the matrix elements
\bea
{\cal M}_{P,11}^2 & = & \l s \frac{h_u^2+h_d^2}{h_uh_d}\, (A_\l + \k
s), \nn\\
{\cal M}_{P,22}^2 & = & 4 \l \k h_u h_d + \l A_\l \frac{h_u h_d}{s}\,
-3 \k A_\k s, \nn\\
{\cal M}_{P,12}^2 & = & \l \sqrt{h_u^2+h_d^2}\, (A_\l - 2\k s).
\eea

\noi It can be diagonalized by an orthogonal $2 \times 2$ matrix
$P_{ij}'$ such that the physical CP-odd states $a_i$ (ordered in mass)
are 
\bea
a_1 &=& P_{11}' \tilde{A} + P_{12}' S_I \nn \\
&=& P_{11}' (\cos\b H_{uI} + \sin\b H_{dI} ) +P_{12}'S_I , \nn \\
a_2 &=& P_{21}' \tilde{A} + P_{22}' S_I \nn \\
&=& P_{21}' (\cos\b H_{uI} + \sin\b H_{dI} ) +P_{22}'S_I , 
\eea

\noi and, for completeness,
\beq
\tilde{G} = -\sin\b H_{uI} + \cos\b H_{dI}\ .
\eeq

\noi The decomposition of the bare states in terms of physical states
reads
\bea
H_{uI} &=& \cos\b (P_{11}' a_1 + P_{21}' a_2) 
- \sin\b \tilde{G} \ ,\nn \\
H_{dI} &=& \sin\b (P_{11}' a_1 + P_{21}' a_2) 
+ \cos\b \tilde{G} \ ,\nn \\
S_I &=& P_{12}' a_1 + P_{22}' a_2\ .
\label{physicaltobare}
\eea

\noi (In principle, since the matrix $P_{ij}'$ is orthogonal, it could
be parameterized by one angle.) Eqs.~(\ref{physicaltobare}) suggest
the introduction of a $2 \times 3$ matrix $P_{ij}$ with
\beq
P_{i1}=\cos\b P_{i1}', P_{i2}=\sin\b P_{i1}', P_{i3}=P_{i2}'
\eeq

\noi such that, omitting the Goldstone boson,
\bea
H_{uI} &=& P_{11} a_1 + P_{21} a_2 \ ,\nn \\
H_{dI} &=& P_{11} a_1 + P_{21} a_2 \ ,\nn \\
S_I &=& P_{13} a_1 + P_{23} a_2\ .
\eea

\subsection{Charged states}

In the basis $(H_u^+, [H_d^-]^* = H^+_d)$, the charged Higgs mass
matrix is given by
\beq
{\cal M}_\pm^2 = \left(\l s (A_\l + \k s) + h_u h_d
(\frac{g_2^2}{2} - \l^2)\right)
\left(\ba{cc} \cot\b & 1 \\ 1 & \tan\b \ea\right) .
\eeq

\noi This gives one eigenstate $H^\pm$ of mass Tr${\cal M}_\pm^2$ and
one massless goldstone mode $G^\pm$ with
\bea
H_u^\pm &=& \cos\b H^\pm - \sin\b G^\pm\ ,\nn \\
H_d^\pm &=& \sin\b H^\pm + \cos\b G^\pm\ .
\eea

\section{SUSY Particles}

\subsection{Neutralinos}

Denoting the $U(1)_Y$ gaugino by $\l_1$ and the neutral $SU(2)$ gaugino
by $\l_2^3$, the mass terms in the Lagrangian read
\bea
{\cal L} & = & \half M_1 \l_1 \l_1 + \half M_2 \l_2^3 \l_2^3 \nn \\
& & + \l (s \psi_u^0 \psi_d^0 + h_u \psi_d^0 \psi_s + h_d \psi_u^0
\psi_s) - \k s \psi_s \psi_s \nn \\
& & + \frac{i g_1}{\sqrt{2}}\, \l_1 (h_u \psi_u^0 - h_d \psi_d^0) -
\frac{i g_2}{\sqrt{2}}\, \l_2^3 (h_u \psi_u^0 - h_d \psi_d^0) .
\eea

\noi In the basis $\psi^0 = (-i\l_1 , -i\l_2, \psi_u^0, \psi_d^0,
\psi_s)$ one can rewrite
\beq
{\cal L} = - \half (\psi^0)^T {\cal M}_0 (\psi^0) + \mathrm{h.c.}
\eeq

\noi where

\beq
{\cal M}_0 =
\left( \ba{ccccc}
M_1 & 0 & \frac{g_1 h_u}{\sqrt{2}} & -\frac{g_1 h_d}{\sqrt{2}} & 0 \\
& M_2 & -\frac{g_2 h_u}{\sqrt{2}} & \frac{g_2 h_d}{\sqrt{2}} & 0 \\
& & 0 & -\mu & -\l h_d \\
& & & 0 & -\l h_u \\
& & & & 2 \k s
\ea \right) . \eeq

\noi (Recall that here $\mu = \mu_\mathrm{eff} = \l s$). One obtains 5
eigenstates (ordered in mass) $\chi^0_i = N_{ij} \psi^0_j$, with
$N_{ij}$ real, with masses $m_\mathrm{\chi^0_i}$ that are real, but not
necessarily positive.

\subsection{Charginos}

The charged $SU(2)$ gauginos are
$\l^- = \frac{1}{\sqrt{2}}\left(\l_2^1 + i \l_2^2\right)$, 
$\l^+ = \frac{1}{\sqrt{2}}\left(\l_2^1 - i \l_2^2\right)$.

\noi Defining 
\beq
\psi^+ = \left(\ba{c} -i\l^+ \\ \psi_u^+ \ea\right)\ , \qq
\psi^- =  \left(\ba{c} -i\l^- \\ \psi_d^- \ea\right)
\eeq

\noi the Lagrangian can be written as
\beq
{\cal L} = -\half (\psi^+ , \psi^-)
\left(\ba{cc} 0 & X^T \\ X & 0 \ea\right)
\left(\ba{c} \psi^+ \\ \psi^- \ea\right) + \mathrm{h.c.} 
\eeq

\noi with
\beq
X = \left(\ba{cc} M_2 & g_2 h_u \\ g_2 h_d & \mu \ea\right) \ .
\eeq

\noi The mass eigenstates are $\chi^+ = V \psi^+ , \; \chi^- = U
\psi^-$, with
\beq
U = \left(\ba{cc}
\cos\t_U & \sin\t_U \\ -\sin\t_U & \cos\t_U
\ea \right) , \qq 
V = \left(\ba{cc}
\cos\t_V & \sin\t_V \\ -\sin\t_V & \cos\t_V
\ea \right) .
\eeq

\noi Defining
\beq
\gamma = \sqrt{\mathrm{Tr}(X^T X) - 4 \mathrm{det}(X^T X)}
\eeq

\noi one has
\bea
\tan\t_U & = & \frac {g_2^2 (h_d^2-h_u^2) + \mu^2 - M_2^2 - \gamma}
{2g_2 (M_2 h_u + \mu h_d)} , \nn \\
\tan\t_V & = & \frac {g_2^2 (h_u^2-h_d^2) + \mu^2 - M_2^2 - \gamma}
{2g_2 (M_2 h_d + \mu h_u)}
\eea

\noi where $-\pi/2 \leq \t_U, \t_V \leq \pi/2$ are such that $M_D = U X
V^T$ is diagonal, but not necessarily positive. The masses with
$|m_\mathrm{\wt\chi_1}| < |m_\mathrm{\wt\chi_2}|$ are given by
\bea
m_\mathrm{\wt\chi_{1}} = \cos\t_U(M_2\cos\t_V + g_2h_u\sin\t_V) +
\sin\t_U(g_2h_d\cos\t_V + \mu\sin\t_V) ,  \nn \\
m_\mathrm{\wt\chi_{2}} = \sin\t_U(M_2\sin\t_V - g_2h_u\cos\t_V) -
\cos\t_U(g_2h_d\sin\t_V - \mu\cos\t_V) . \nn \\
\eea

\noi In terms of 4 component
Dirac spinors $\Psi_i = \left(\ba{c} \chi^+_i \\ \overline{\chi}^{\
-}_i \ea\right)$ one can rewrite the Lagrangian as
\beq
{\cal L} = - \chi^- M_D \chi^+ + \mathrm{h.c.} = - m_\mathrm{\wt\chi_1}
\overline{\Psi}_1 \Psi_1 - m_\mathrm{\wt\chi_2} \overline{\Psi}_2
\Psi_2\ .
\eeq

\subsection{Top and Bottom Squarks}

To complete the consequences of our conventions above, we give here the
top and bottom squark mass-squared matrices (without the D-term
contributions). Below, $t_L$, $t_R^c$, $b_L$ and $b_R^c$ denote the two
component quark spinors.\par

\noi Top squarks:

\beq
\ba{cc}
 & T_R \hskip 2cm T_L  \\
\ba{c} T_R^*  \\ T_L^*  \ea &
\left(\ba{cc} m_T^2 + h_t^2 h_u^2 & h_t (A_t h_u - \l s h_d)  \\
 h_t (A_t h_u - \l s h_d) & m_Q^2 + h_t^2 h_u^2 \ea \right)
\ea
\eeq
 
\noi Bottom squarks:

\beq
\ba{cc}
 & B_R \hskip 2cm B_L  \\
\ba{c} B_R^*  \\ B_L^*  \ea &
\left(\ba{cc} m_B^2 + h_b^2 h_d^2 & h_b  (A_b h_d-\l s h_u) \\
h_b (A_b h_d-\l s h_u) & m_Q^2 + h_b^2 h_d^2 \ea \right)
\ea
\eeq

\newpage
\setcounter{equation}{0}
\setcounter{section}{0}
\renewcommand{\theequation}{B.\arabic{equation}}

\section*{Appendix B} 

\section{Feynman rules for the Higgs Couplings}


\subsection{Higgs-Quarks}

The couplings are obtained by expanding the quark mass matrices in the
(properly normalized) physical Higgs fields $h_i$, $a_i$ and $H^\pm$.
Below, we use $v^2 = h_u^2 + h_d^2$, and consider the quarks of
the third generation.
\bea
h_i t_L t_R^c & : & \frac{m_t}{\sqrt{2}v\sin\b} S_{i1} \nn \\
h_i b_L b_R^c & : & \frac{m_b}{\sqrt{2}v\cos\b} S_{i2} \nn \\
a_i t_L t_R^c & : & i\frac{m_t}{\sqrt{2}v\sin\b} P_{i1} \nn \\
a_i b_L b_R^c & : & i\frac{m_b}{\sqrt{2}v\cos\b} P_{i2} \nn \\
H^+ b_L t_R^c & : & -\frac{m_t}{v} \cot\b \nn \\
H^- t_L b_R^c & : & -\frac{m_b}{v} \tan\b
\eea

\subsection{Higgs-Gauge Bosons}

\noi These couplings are obtained from the kinetic terms in the
Lagrangian:  
\bea
h_i Z_\mu Z_\nu & : & g_{\mu\nu} \frac{g_1^2 + g_2^2}{\sqrt{2}}
(h_u S_{i1} + h_d S_{i2})\nn \\
h_i W^+_\mu W^-_\nu & : & g_{\mu \nu} \frac{g_2^2}{\sqrt{2}} 
(h_u S_{i1} + h_d S_{i2}) \nn \\
h_i(p) H^+(p') W_\mu^- & : & \frac{g_2}{2} (\cos\b S_{i1} - 
\sin\b S_{i2}) (p - p')_\mu \nn \\
a_i(p) H^+(p') W_\mu^- & : & i\frac{g_2}{2} (\cos\b P_{i1} + 
\sin\b P_{i2}) (p - p')_\mu \nn \\
h_i(p) a_j(p') Z_\mu & : & i\frac{g}{\sqrt{2}} (S_{i1} P_{j1} - 
S_{i2} P_{j2}) (p - p')_\mu \nn \\
H^+(p) H^-(p') Z_\mu & : & \frac{g_1^2-g_2^2}{\sqrt{g_1^2+g_2^2}}
(p - p')_\mu
\eea

\subsection{Higgs-Neutralinos/Charginos}

As in the case of the Higgs-Quark couplings, these couplings are
obtained by expanding the corresponding mass matrices:

\bea
h_a \chi^+_i \chi^-_j & : & \frac{\l}{\sqrt{2}} S_{a3} U_{i2} V_{j2} +
\frac{g_2}{\sqrt{2}} (S_{a1} U_{i1} V_{j2} + S_{a2} U_{i2} V_{j1}) \nn
\\
a_a \chi^+_i \chi^-_j & : & i\left(\frac{\l}{\sqrt{2}} P_{a3} U_{i2}
V_{j2} - \frac{g_2}{\sqrt{2}} (P_{a1} U_{i1} V_{j2} + P_{a2} U_{i2}
V_{j1})\right) \nn \\
H^+ \chi^-_i \chi^0_j & : & \l\cos\b U_{i2} N_{j5} -
\frac{\sin\b}{\sqrt{2}} U_{i2} (g_1 N_{j1} + g_2 N_{j2}) + g_2 \sin\b
U_{i1} N_{j4} \nn \\
H^- \chi^+_i \chi^0_j & : & \l\sin\b V_{i2} N_{j5} +
\frac{\cos\b}{\sqrt{2}} V_{i2} (g_1 N_{j1} + g_2 N_{j2}) + g_2 \cos\b
V_{i1} N_{j3} \nn \\
h_a \chi^0_i \chi^0_j & : & \frac{\l}{\sqrt{2}} (S_{a1} \Pi_{ij}^{45} +
S_{a2} \Pi_{ij}^{35} + S_{a3} \Pi_{ij}^{34}) - \sqrt{2} \k S_{a3}
N_{i5} N_{j5} \nn \\ & & - \frac{g_1}{2} (S_{a1} \Pi_{ij}^{13} - S_{a2}
\Pi_{ij}^{14}) + \frac{g_2}{2} (S_{a1} \Pi_{ij}^{23} - S_{a2}
\Pi_{ij}^{24}) \nn \\
a_a \chi^0_i \chi^0_j & : & i\left(\frac{\l}{\sqrt{2}} (P_{a1}
\Pi_{ij}^{45} + P_{a2} \Pi_{ij}^{35} + P_{a3} \Pi_{ij}^{34}) - \sqrt{2}
\k P_{a3} N_{i5} N_{j5}\right. \nn \\ 
& & \left. + \frac{g_1}{2} (P_{a1} \Pi_{ij}^{13} - P_{a2}
\Pi_{ij}^{14}) - \frac{g_2}{2} (P_{a1} \Pi_{ij}^{23} - P_{a2}
\Pi_{ij}^{24})\right)
\eea

\noi where $\Pi_{ij}^{ab} = N_{ia}N_{jb}+N_{ib}N_{ja}$.

\subsection{Triple Higgs Interactions}

\noi The trilinear Higgs self-couplings are obtained by expanding the scalar
potential.
\bea
h_a h_b h_c & : &
\frac{\l^2}{\sqrt{2}} \left( h_u (\Pi_{abc}^{122}+\Pi_{abc}^{133}) +
h_d (\Pi_{abc}^{211}+\Pi_{abc}^{233}) + s
(\Pi_{abc}^{311}+\Pi_{abc}^{322}) \right) \nn \\
& & - \frac{\l\k}{\sqrt{2}} ( h_u \Pi_{abc}^{323} + h_d \Pi_{abc}^{313}
+ 2 s \Pi_{abc}^{123}) + \sqrt{2} \k^2 s \Pi_{abc}^{333} \nn \\
& & - \frac{\l A_\l}{\sqrt{2}} \Pi_{abc}^{123} + \frac{\k
A_\k}{3\sqrt{2}} \Pi_{abc}^{333} \nn \\
& & + \frac{g^2}{2\sqrt{2}} \left( h_u (\Pi_{abc}^{111} -
\Pi_{abc}^{122}) - h_d (\Pi_{abc}^{211} - \Pi_{abc}^{222}) \right)
\eea

\noi where
\bea
\Pi_{abc}^{ijk} & = &
S_{ai} S_{bj} S_{ck} + S_{ai} S_{cj} S_{bk} + S_{bi} S_{aj} S_{ck} \nn
\\ & & + S_{bi} S_{cj} S_{ak} + S_{ci} S_{aj} S_{bk} + S_{ci} S_{bj}
S_{ak}\ .
\eea

\bea
h_a a_b a_c & : &
\frac{\l^2}{\sqrt{2}} \left( h_u (\Pi_{abc}^{122}+\Pi_{abc}^{133}) +
h_d (\Pi_{abc}^{211}+\Pi_{abc}^{233}) + s
(\Pi_{abc}^{311}+\Pi_{abc}^{322}) \right) \nn \\
& & + \frac{\l\k}{\sqrt{2}} \left( h_u
(\Pi_{abc}^{233}-2\Pi_{abc}^{323}) + h_d
(\Pi_{abc}^{133}-2\Pi_{abc}^{313}) \right. \nn \\
& & \left. + 2 s (\Pi_{abc}^{312}-\Pi_{abc}^{123}-\Pi_{abc}^{213})
\right) + \sqrt{2} \k^2 s \Pi_{abc}^{333} \nn \\
& & + \frac{\l A_\l}{\sqrt{2}}
(\Pi_{abc}^{123}+\Pi_{abc}^{213}+\Pi_{abc}^{312}) - \frac{\k
A_\k}{\sqrt{2}} \Pi_{abc}^{333} \nn \\
& & + \frac{g^2}{2\sqrt{2}} \left( h_u (\Pi_{abc}^{111} -
\Pi_{abc}^{122}) - h_d (\Pi_{abc}^{211} - \Pi_{abc}^{222}) \right)
\eea

\noi where
\bea
\Pi_{abc}^{ijk} = S_{ai} (P_{bj} P_{ck} + P_{cj} P_{bk})\ .
\eea

\bea
h_a H^+ H^- & : &  \frac{\l^2}{\sqrt{2}} ( s (\Pi_a^{311}+\Pi_a^{322})
- h_u \Pi_a^{212} - h_d \Pi_a^{112}) \nn \\
& & + \sqrt{2} \l \k s \Pi_a^{312} + \frac{\l A_\l}{\sqrt{2}}
\Pi_a^{312} \nn \\
& & + \frac{g_1^2}{4\sqrt{2}} \left( h_u (\Pi_a^{111}-\Pi_a^{122}) + 
h_d (\Pi_a^{222}-\Pi_a^{211}) \right) \\
& & + \frac{g_2^2}{4\sqrt{2}} \left( h_u
(\Pi_a^{111}+\Pi_a^{122}+2\Pi_a^{212}) + h_d
(\Pi_a^{211}+\Pi_a^{222}+2\Pi_a^{112}) \right) \nn
\eea

\noi where
\bea
\Pi_{a}^{ijk} = 2 S_{ai} C_j C_k
\eea

\noi with $C_1 = \cos\b$, $C_2 = \sin\b$.
\newpage

\setcounter{equation}{0}
\setcounter{section}{0}
\renewcommand{\theequation}{C.\arabic{equation}}

\section*{Appendix C} 

In this appendix we give the explicit expressions of the radiative
corrections to the Higgs masses.

Generally, the radiative corrections to the Higgs effective action
consist in corrections to the wave function renormalization constants
$Z_i$, and corrections to the effective potential $V_{eff}(h_i)$. We
will treat the one loop corrections originating from stop and sbottom
loops exactly in the stop and sbottom mixings and mass splittings. The
resulting contributions to the Higgs masses are nevertheless quite
simple if they are expressed in terms of couplings and masses at the
scale $Q=M_{SUSY}$ (the average of the squark masses, see below) and in
terms of the Higgs vevs {\it before} rescaling by $Z_i^{1/2}$, which we
will denote by $h_i(Q)$. However, the Higgs vevs are related to the
Fermi coupling $G_F$
and our definition of $\tan{\beta}$ {\it after} rescaling by
$Z_i^{1/2}$, cf. eq. (2.9). 

Hence, we first have to determine $Z_i$ for $i = H_u, H_d$:
\bea
Z_{H_u} &=& 1 + \frac{3 h_t^2}{16 \pi^2} t\nn \\
Z_{H_d} &=& 1 + \frac{3 h_b^2}{16 \pi^2} t
\eea
\noi where
\bea
t &=& \ln(Q^2/m_{top}^2)\ ,\nn \\
Q^2 &=& \frac{1}{4}(2 m_{Q}^2 + m_{T}^2 + m_{B}^2) + m_{top}^2
\eea
\noi (such that $t=0$ for vanishing SUSY breaking squark masses
$m_{Q}$, $m_{T}$ and $m_{B}$). Now we have
\beq
h_u(Q)=h_u/\sqrt{Z_{H_u}}\ ,\qquad h_d(Q)=h_d/\sqrt{Z_{H_u}}\ .
\eeq
\noi with $h_u$ and $h_d$ given in terms of $G_F$ and $\tan{\beta}$.

In addition we need the running Yukawa couplings and quark masses at
the scale $Q$. First, the running bottom quark mass (and hence $h_b$)
at the scale $m_{top}$ is given by its QCD evolution to one loop order
\beq
h_b(m_{top}) = h_b(m_{bot})\left( 1-\frac{23}{12\pi}\alpha_s(m_{top})
\ln{\frac{m_{top}^2}{m_{bot}^2}}\right) ^{12/23}\ .
\eeq
\noi The running top quark Yukawa coupling at the scale $m_{top}$ is
given in terms of the top quark pole mass as \footnote{Here we use the
$\overline{MS}$ relation between the pole mass and the running mass. In
the $\overline{DR}$ scheme the second factor $4/3\pi$ would read
$5/3\pi$, which would increase $h_t(m_{top})$ and hence the lightest
Higgs mass by $\sim 1\%$. Since we have not included subdominant
(single) logarithms in the two loop corrections to the effective
potential we are not sensitive to the scheme in which the running top
quark mass is defined, which leads to a theoretical error of $\sim 1\%$
on the mass of the lightest Higgs.}
\beq
h_t(m_{top}) = \frac{m_{top}(pole)}{h_u}\left( 1+
\frac{4\alpha_s(m_{top})}{3\pi}
+\frac{11\alpha_s^2(m_{top})}{\pi^2}\right)\ .
\eeq

Next, the running Yukawa couplings at the scale $Q$ are obtained from
the RG equations (for an effective 2 Higgs doublet model)
\bea
\frac{dh_t}{d\ln{Q^2}} &=& \frac{h_t}{64\pi^2}(9h_t^2 + h_b^2)
-h_t\frac{\alpha_s}{\pi}\ ,\nn \\
\frac{dh_b}{d\ln{Q^2}} &=& \frac{h_b}{64\pi^2}(9h_b^2 + h_t^2)
-h_b\frac{\alpha_s}{\pi}\ .
\eea

Its solutions, exact in $\alpha_s$ but perturbative in the Yukawa
couplings, are
\bea
h_t(Q) &=&
h_t(m_{top})\left(1+\frac{7}{4\pi}\alpha_s(m_{top})t\right)^{-4/7}
\left( 1+\frac{1}{64\pi^2}(9h_t^2+h_b^2)t\right) \ ,\nn \\
h_b(Q) &=&
h_b(m_{top})\left(1+\frac{7}{4\pi}\alpha_s(m_{top})t\right)^{-4/7}
\left( 1+\frac{1}{64\pi^2}(9h_b^2+h_t^2)t\right) \ .
\eea

However, possibly a $SU(2)$ Higgs doublet has a mass much larger than
$m_{top}$. Then it is practically degenerate, and its approximate masses
are given in the NMSSM by
\beq
M_{HD} =
\lambda s (A_{\lambda}+\kappa s)
(\cot{\beta}+\tan{\beta})\ .
\eeq
\noi At scales below $M_{HD}$ the RG equations read
\bea
\frac{dh_t}{d\ln{Q^2}} &=& \frac{h_t}{64\pi^2}(9h_t^2 + 3h_b^2)
-h_t\frac{\alpha_s}{\pi}\ ,\nn \\
\frac{dh_b}{d\ln{Q^2}} &=& \frac{h_b}{64\pi^2}(9h_b^2 + 3h_t^2)
-h_b\frac{\alpha_s}{\pi}\ .
\eea
In order to correct the expressions for $h_t(Q)$, $h_b(Q)$ in this
case, we have to multiply them by $\left(1+\frac{h_b^2}{32\pi^2}
\ln({M_{HD}^2/m_{top}^2})\right)$, $\left(1+\frac{h_t^2}{32\pi^2}
\ln({M_{HD}^2/m_{top}^2})\right)$, respectively, if $M_{HD} > m_{top}$.

The expressions for the running top and bottom quark masses at the scale
$Q$ are then simply
\beq
m_{top}(Q)= h_t(Q) h_u(Q)\ , \qquad m_{bot}(Q)= h_b(Q) h_d(Q)\ ,
\eeq
and below it is convenient to use 
\beq
\tan{\beta}(Q) = h_u(Q)/h_d(Q)\ .
\eeq

For the (s)top, (s)bottom induced one loop corrections to $V_{eff}$
we still need the eigenvalues of the stop, sbottom mass matrices (A.32)
and (A.33). To this end it is convenient to define, step by step,
\bea
M_{stop}^2 &=& \frac{1}{2}(m_{Q}^2 + m_{T}^2)\ ,\nn \\
M_{sbot}^2 &=& \frac{1}{2}(m_{Q}^2 + m_{B}^2)\ ,\nn \\
\Delta M_{stop}^2 &=& \frac{1}{2}(m_{Q}^2 - m_{T}^2)\ ,\nn \\
\Delta M_{sbot}^2 &=& \frac{1}{2}(m_{Q}^2 - m_{B}^2)\ ,\nn \\
X_t &=& A_{t} - \lambda s\cot{\beta}(Q)\ ,\nn \\
X_b &=& A_{b} - \lambda s\tan{\beta}(Q)\ ,\nn \\
W_t &=& \sqrt{\Delta M_{stop}^4 +m_{top}^2(Q) X_t^2}\ ,\nn \\
W_b &=& \sqrt{\Delta M_{sbot}^4 +m_{bot}^2(Q) X_b^2}\ ,
\eea
with the help of which the eigenvalues of the stop, sbottom mass 
matrices become
\bea
M_{stop1}^2 &=& M_{stop}^2 + m_{top}^2(Q) - W_t\ ,\nn \\
M_{stop2}^2 &=& M_{stop}^2 + m_{top}^2(Q) + W_t\ ,\nn \\
M_{sbot1}^2 &=& M_{sbot}^2 + m_{bot}^2(Q) - W_b\ ,\nn \\
M_{sbot2}^2 &=& M_{sbot}^2 + m_{bot}^2(Q) + W_b\ .
\eea
Finally the following auxiliary quantities are useful:
\bea
fmt &=& \frac{1}{2W_t}\left( M_{stop2}^2
\ln{\left(\frac{M_{stop2}^2}{Q^2}\right)} - 
M_{stop1}^2\ln{\left(\frac{M_{stop1}^2}{Q^2}\right)}\right) -1\ ,\nn \\
gmt &=& \frac{m_{top}^2 X_t^2}{W_t^2}
\left(\frac{M_{stop2}^2 + M_{stop1}^2}{M_{stop2}^2 - M_{stop1}^2}
\ln{\left(\frac{M_{stop2}^2}{M_{stop1}^2}\right)} -2\right)\ ,\nn \\
emt &=& \frac{m_{top}^2 X_t}{W_t}
\ln{\left(\frac{M_{stop2}^2}{M_{stop1}^2}\right)}\ ,\nn \\
fmb &=& \frac{1}{2W_b}\left( M_{sbot2}^2
\ln{\left(\frac{M_{sbot2}^2}{Q^2}\right)} - 
M_{sbot1}^2\ln{\left(\frac{M_{sbot1}^2}{Q^2}\right)}\right)  -1\ ,\nn \\
gmb &=& \frac{m_{bot}^2 X_b^2}{W_b^2}
\left(\frac{M_{sbot2}^2 + M_{sbot1}^2}{M_{sbot2}^2 - M_{sbot1}^2}
\ln{\left(\frac{M_{sbot2}^2}{M_{sbot1}^2}\right)} -2\right)\ ,\nn \\
emb &=& \frac{m_{bot}^2 X_b}{W_b}
\ln{\left(\frac{M_{sbot2}^2}{M_{sbot1}^2}\right)}\ ,
\eea
and, remarkably, quite many of the radiative corrections can be absorbed
into a shift of $A_{\lambda}$:
\beq
A_{\lambda,new} = A_{\lambda} + \frac{3 h_t(Q)^2}{16\pi^2} A_t\ fmt +
\frac{3 h_b(Q)^2}{16\pi^2} A_b\ fmb\ .
\eeq
Subsequently the index $new$ of $A_{\lambda,new}$ will be omitted.

Then the radiatively corrected mass-squared matrix entries in the CP
even sector are, instead of (A.9),

\bea
{\cal M}_{S,11}^2 & = & g^2 h_u(Q)^2 + \l s  (A_\l + \k s)
\cot{\beta}(Q) 
- \frac{3 h_b(Q)^2}{32\pi^2}(\lambda s)^2\ gmb
\nn\\
&&+ \frac{3 h_t(Q)^2}{32\pi^2}\left(-A_t^2\ gmt + 4A_t\ emt
+ 4 m_{top}^2\ln\left({\frac{M_{stop1}^2 M_{stop2}^2}{m_{top}^4}}\right)
\right)\ , \nn\\
{\cal M}_{S,22}^2 & = & g^2 h_d(Q)^2 + \l s (A_\l + \k s)
\tan{\beta}(Q) - \frac{3 h_t(Q)^2}{32\pi^2}(\lambda s)^2\ gmt
\nn\\
&&+ \frac{3 h_b(Q)^2}{32\pi^2}\left(-A_b^2\ gmt + 4A_b\ emb
+ 4 m_{bot}^2\ln\left({\frac{M_{sbot1}^2 M_{sbot2}^2}{m_{bot}^4}}\right)
\right)\ , \nn\\
{\cal M}_{S,33}^2 & = & \l A_\l \frac{h_u(Q) h_d(Q)}{s}\, + \k s (A_\k
+ 4 \k s)\nn\\
&&- \frac{3 h_t(Q)^2}{32\pi^2}\lambda^2h_d(Q)^2\ gmt
- \frac{3 h_b(Q)^2}{32\pi^2}\lambda^2h_u(Q)^2\ gmb
\ , \nn\\
{\cal M}_{S,12}^2 & = & (2\l^2 - g^2) h_u(Q) h_d(Q) - \l s (A_\l + \k s)
\nn\\
&&+ \lambda s \left(\frac{3 h_t(Q)^2}{32\pi^2}(A_t\ gmt -2\ emt)
+ \frac{3 h_b(Q)^2}{32\pi^2}(A_b\ gmb -2\ emb)\right)
\ , \nn\\ 
{\cal M}_{S,13}^2 & = & 2\l^2 h_u(Q) s - \l h_d(Q) (A_\l + 2\k s)
\nn\\
&&+  \frac{3 h_t(Q)^2}{32\pi^2}\lambda h_d(Q) (A_t\ gmt-2\ emt)\nn\\
&&+\frac{3 h_b(Q)^2}{32\pi^2}\lambda^2 s\ h_u(Q)(4\ fmb-gmb)
\ ,\nn\\
{\cal M}_{S,23}^2 & = & 2\l^2 h_d(Q) s - \l h_u(Q) (A_\l + 2\k s)
\nn\\
&&+\frac{3 h_b(Q)^2}{32\pi^2} \lambda h_u(Q) (A_b\ gmb-2\ emb)\nn\\
&&+\frac{3 h_t(Q)^2}{32\pi^2}\lambda^2 s\ h_d(Q)(4\ fmt-gmt) 
\ .
\eea
The dominant two loop effects $\sim h_t^6 t^2$ and $\sim
h_t^4\alpha_s t^2$ contribute to ${\cal M}_{S,11}^2$ only. They read
\beq
\delta{\cal M}_{S,11\ 2loops}^2 = \frac{3 h_t(Q)^4}{64\pi^4}h_u(Q)^2
\left( 32\pi\alpha_s(Q)-\frac{3}{2}h_t(Q)^2\right) t^2\ .
\eeq
The above expressions for the elements of the CP even mass matrix have
been obtained by taking the second derivatives of the effective
potential, and using the three minimization equations with respect to
$h_u$, $h_d$ and $s$ in order to eliminate the three SUSY breaking
masses squared. Since the Higgs fields are not quite properly
normalized due to the $Z$ factors, the above mass matrix elements still
have to be rescaled by appropriate powers of $Z_i$ 
(here below ${\cal M}_{S,ij\ norm.}^2$ denote the mass matrix elements
for correctly normalized Higgs fields):
\bea
{\cal M}_{S,11\ norm.}^2 & = & {\cal M}_{S,11}^2 / Z_{H_u}\ ,\nn\\
{\cal M}_{S,22\ norm.}^2 & = & {\cal M}_{S,22}^2 / Z_{H_d}\ ,\nn\\
{\cal M}_{S,12\ norm.}^2 & = & {\cal M}_{S,12}^2 / \sqrt{Z_{H_u}Z_{H_d}}
\ ,\nn\\
{\cal M}_{S,13\ norm.}^2 & = & {\cal M}_{S,13}^2 /\sqrt{Z_{H_u}}
\ ,\nn\\
{\cal M}_{S,23\ norm.}^2 & = & {\cal M}_{S,23}^2 / \sqrt{Z_{H_d}}
\ .
\eea

Now we turn to the electroweak leading logarithmic corrections to the
Higgs mass matrix elements, which are taken from ref. [15]. As in [15]
we allow for a Higgs doublet which is possibly heavy ($M_{HD}\ >\
M_Z$), and for gauginos/Higgsinos with masses possibly (much) larger
than $M_Z$. In this case their masses are given approximately by the
soft SUSY breaking gaugino mass parameters $M_1$, $M_2$ or $\mu_{eff} =
\lambda s$, respectively. The following auxiliary parameters are useful
subsequently:
\bea
\mu_1 &=& max\{|\mu_{eff}|,\ M_1\}\ ,\nn\\
\mu_2 &=& max\{|\mu_{eff}|,\ M_2\}\ ,\nn\\
\mu_{12} &=& max\{|\mu_{eff}|,\ M_1,\ M_2\}\ .
\eea

The electroweak corrections to the Higgs mass matrix elements are then
given by (we omit the index $norm.$ in the following)

\bea
\delta_{ew}{\cal M}_{S,11}^2 &=& \frac{g_2^2 M_Z^4}{16\pi^2 M_W^2}
\sin^2{\beta} \left( (4-8\sin^2\theta_W+\frac{32}{3} \sin^4\theta_W) t
\right. \nn\\
&& \left. 
+\frac{1}{6}[-9(\cos^2{\beta}-\sin^2{\beta})^4 +(1-2\sin^2{\theta_W}
+2\sin^4{\theta_W})(\cos^2{\beta}-\sin^2{\beta})^2\right. \nn\\
&& \left.
+10-2\sin^2\theta_W
+2\sin^4\theta_W]\ln(M_{HD}^2/M_Z^2)\right.\nn \\
&&\left.  + \sin^2\theta_W(1-2\sin^2\theta_W)\ln(\mu_1^2/M_Z^2)
-4\sin^2\theta_W\cos^2\theta_W\ln(\mu_{12}^2/M_Z^2)\right.\nn \\
&&\left.  +\cos^2\theta_W(3-10\cos^2\theta_W)\ln(\mu_2^2/M_Z^2)-
\frac{4}{3}\cos^4\theta_W\ln(M_2^2/M_Z^2)\right. \nn\\
&& \left.
-\frac{2}{3}(\sin^2\theta_W+\cos^4\theta_W)\ln(\mu_{eff}^2/M_Z^2) 
\right)\ ,\nn\\
\delta_{ew}{\cal M}_{S,22}^2 &=& \frac{g_2^2 M_Z^4}{16\pi^2 M_W^2}
\cos^2{\beta} \left( (4-8\sin^2\theta_W+\frac{32}{3}\sin^4\theta_W) t
\right. \nn\\
&& \left. +\frac{1}{6}[-9(\cos^2{\beta}-\sin^2{\beta})^4
+(1-2\sin^2{\theta_W}+2\sin^4{\theta_W})
(\cos^2{\beta}-\sin^2{\beta})^2\right. \nn\\
&& \left. +10-2\sin^2\theta_W
+2\sin^4\theta_W]\ln(M_{HD}^2/M_Z^2)\right.\nn \\
&&\left. + \sin^2\theta_W(1-2\sin^2\theta_W)\ln(\mu_1^2/M_Z^2)
-4\sin^2\theta_W\cos^2\theta_W\ln(\mu_{12}^2/M_Z^2)\right.\nn \\
&&\left.  +\cos^2\theta_W(3-10\cos^2\theta_W)\ln(\mu_2^2/M_Z^2)-
\frac{4}{3}\cos^4\theta_W\ln(M_2^2/M_Z^2)\right. \nn\\
&& \left.
-\frac{2}{3}(\sin^2\theta_W+\cos^4\theta_W)\ln(\mu_{eff}^2/M_Z^2) 
\right)\ ,\nn\\
\delta_{ew}{\cal M}_{S,12}^2 &=& \frac{g_2^2 M_Z^4}{16\pi^2 M_W^2}
\sin\beta \cos{\beta} 
\left( (-4+8\sin^2\theta_W-\frac{32}{3}\sin^4\theta_W) t
\right. \nn\\
&& \left. -\frac{1}{6}[-9(\cos^2{\beta}-\sin^2{\beta})^4
+(1-2\sin^2{\theta_W}+2\sin^4{\theta_W})
(\cos^2{\beta}-\sin^2{\beta})^2\right. \nn\\
&& \left. +8-22\sin^2\theta_W
+10\sin^4\theta_W]\ln(M_{HD}^2/M_Z^2)\right.\nn \\
&&\left. - \sin^2\theta_W(1+2\sin^2\theta_W)\ln(\mu_1^2/M_Z^2)
-4\sin^2\theta_W\cos^2\theta_W\ln(\mu_{12}^2/M_Z^2)\right.\nn \\
&&\left.  -\cos^2\theta_W(3+2\cos^2\theta_W)\ln(\mu_2^2/M_Z^2)+
\frac{4}{3}\cos^4\theta_W\ln(M_2^2/M_Z^2)\right. \nn\\
&& \left.
+\frac{2}{3}(\sin^2\theta_W+\cos^4\theta_W)\ln(\mu_{eff}^2/M_Z^2) 
\right)\ .
\eea

The eigenvalues of this Higgs mass matrix give the "running" Higgs
masses at a scale $m_{top}$ due to the argument of the logarithm t in
the $Z$ factors (C.1). In the LLA we can identify the scales $m_{top}$,
$M_Z$ (in the arguments of the logarithms) and, to start with,
$M_{Higgs}$, and take these "running" masses as pole masses. However,
if a Higgs mass is (much) larger than $m_{top}$, we should have
computed the $Z$ factors (C.1) at a scale $M_{Higgs}$ rather than
$m_{top}$, in order to identify the eigenvalues of the Higgs mass
matrix with the Higgs pole masses. We can correct for this at the end
(after having diagonalized the Higgs mass matrix), taking care that now
the eigenstates $h_i$ are given by $h_i = S_{ij}S_j^{bare}$, cf. the
notation below eq. (A.9). Hence, if $M_{Higgs} > m_{top}$, the Higgs
pole masses $M_{i}^2(pole)$ are given in terms of the eigenvalues
$M_i^2$ as
\beq
M_{i}^2(pole) = M_i^2/\left(1-(\frac{3 h_t^2}{16\pi^2}S_{i1}^2 
+\frac{3 h_b^2}{16\pi^2}S_{i2}^2)\ln{(M_i^2/m_{top}^2)}\right)\ .
\eeq

Now we turn to the radiative corrections to the CP odd mass matrix.
After the shift (C.15) in $A_{\lambda}$ these are trivial: It suffices
to take the tree level expressions (A.11) for the mass matrix elements,
to replace the Higgs vevs $h_u$ and $h_d$ by its expressions (C.3)
at the scale $Q$, and to rescale the mass matrix elements by appropriate
powers of $Z_i$ as in (C.18) for the CP even sector. After dropping
the Goldstone mode this gives a $2\ \times\ 2$ mass matrix (instead
of (A.13))
\bea
{\cal M}_{P,11}^2 & = & \l s (A_\l + \k s)
\left( \frac{\tan{\beta}(Q)}{ Z_{H_d}}
+\frac{\cot{\beta}(Q)}{Z_{H_u}}\right)\ , \nn\\
{\cal M}_{P,22}^2 & = & 4 \l \k h_u(Q) h_d(Q) + \l A_\l 
\frac{h_u(Q) h_d(Q)}{s}\, -3 \k A_\k s\ , \nn\\
{\cal M}_{P,12}^2 & = & \l \sqrt{h_u(Q)^2/Z_{H_d}+h_d(Q)^2/Z_{H_u}}
\, (A_\l - 2\k s)\ .
\eea
Finally, if a CP odd Higgs mass is (much) larger than $m_{top}$, its
pole mass is given by an expression analogous to (C.21) after replacing
$S_{i1}$ by $P'_{i1}\cos^2{\beta}$ and $S_{i2}$ by
$P'_{i1}\sin^2{\beta}$ (see (A.14) for the definition of $P'_{ij}$).

In the expression for radiatively corrected charged Higgs mass we
neglect stop/\-sbottom loops (without large logarithms), but we include
large logarithms proportional to electroweak gauge couplings:
\bea
{M}_{H^\pm}^2 & = & \left( \lambda s (A_\lambda + \k s) + h_u(Q)
h_d(Q)(g_2^2/2-\lambda^2)\right)\left(\frac{\tan{\beta}(Q)}{Z_{H_d}} +
\frac{\cot{\beta}(Q)}{Z_{H_u}}\right)\nn\\
&&+ 3 h_t^2 h_b^2 t/(16\sqrt{2}\pi^2 G_F) \nn\\
&&+\frac{g_2^2 M_W^2}{48 \pi^2}\left(12 t +3 \tan^2\theta_W
(\ln(\mu_1^2/M_Z^2) + 4 \ln(\mu_{12}^2/M_Z^2)) \right. \nn\\
&&\left. -3\ln(\mu_2^2/M_Z^2) -4\ln(M_2^2/M_Z^2)
-2\ln(\mu_{eff}^2/M_Z^2)\right)
\eea
This time, once ${M}_{H^\pm}$ is larger than $m_{top}$, its pole
mass is given by
\beq
{M}_{H^\pm}^2 (pole) = {M}_{H^\pm}^2 /\left(1-
\left(\frac{3 h_t^2}{16\pi^2}\cos^2{\beta}
+\frac{3 h_b^2}{16\pi^2}\sin^2{\beta}\right)\ln{(M_{H^\pm}^2/m_{top}^2)}
\right)\ .
\eeq

\end{document}